\documentclass[prl,twocolumn,preprintnumbers,amsmath,amssymb]{revtex4}

\usepackage{graphicx}
\usepackage{dcolumn}
\usepackage{bm}
\usepackage{color}
\usepackage{epstopdf}

\usepackage{ulem}

\def\bk{{\bm \kappa}}

\newcommand{\br}{ {\bm r}}

\def\EE{{\mathbb{E}}}

\def\bsigma{{\boldsymbol{\sigma}}}

\def\bk{{\bm k}}

\def\br{{\bm r}}

\begin{document}



\title{Observation of light thermalization to negative temperature\\ Rayleigh-Jeans equilibrium states in multimode optical fibers}
\author{K. Baudin$^{1,2}$, J. Garnier$^{2}$, A. Fusaro$^{3}$, N. Berti$^{1}$, C. Michel$^{4}$,  K. Krupa$^{5}$, G. Millot$^{1,6}$, A. Picozzi$^{1}$}
\affiliation{$^{1}$ Laboratoire Interdisciplinaire Carnot de Bourgogne, CNRS, Universit\'e Bourgogne Franche-Comt\'e, Dijon, France}
\affiliation{$^{2}$ CMAP, CNRS, Ecole Polytechnique, Institut Polytechnique de Paris, 91128 Palaiseau Cedex, France}
\affiliation{$^{3}$ CEA, DAM, DIF, F-91297 Arpajon Cedex, France} 
\affiliation{$^{4}$ Universit\'e C\^ote d'Azur, CNRS, Institut de Physique de Nice, Nice, France}
\affiliation{$^{5}$ Institute of Physical Chemistry Polish Academy of Sciences, Warsaw, Poland}
\affiliation{$^{6}$ Institut Universitaire de France (IUF), 1 rue Descartes, 75005 Paris, France}


\begin{abstract}
Although the temperature of a thermodynamic system is usually believed to be a positive quantity, under particular conditions, negative temperature equilibrium states are also possible.
Negative temperature equilibriums have been observed with spin systems, cold atoms in optical lattices and two-dimensional quantum superfluids.
Here we report the observation of Rayleigh-Jeans thermalization of light waves to negative temperature equilibrium states.
The optical wave relaxes to the equilibrium state through its propagation in a multimode optical fiber, i.e., in a conservative Hamiltonian system.
The bounded energy spectrum of the optical fiber enables negative temperature equilibriums with high energy levels (high order fiber modes) more populated than low energy levels (low order modes).
Our experiments show that negative temperature speckle beams are featured, in average, by a non-monotonous radial intensity profile.
The experimental results are in quantitative agreement with the Rayleigh-Jeans theory without free parameters.
Bringing negative temperatures to the field of optics opens the door to the investigation of fundamental issues of negative temperature states in a flexible experimental environment.
\end{abstract}


\maketitle

{\it Introduction.-} 
Temperature is a central concept of statistical mechanics and often reflects a measure of the amount of disordered motion in a classical ideal gas.
Although this intuitive notion is correct for many physical systems, one should keep in mind that the concept of temperature is by far more subtle.
A detailed analysis of the concept of temperature, and of its relationship with energy and entropy shows that, under suitable conditions, the entropy can decrease with the energy, thus allowing for the existence of equilibrium states at negative temperatures (NT). 
Starting from the seminal works by Onsager \cite{onsager49} and Ramsey \cite{ramsey56}, who originally conceived the physical idea and the first theoretical approaches, during the last decades, many works have been devoted to the theoretical understanding of these unusual equilibrium states. 
Despite the fact that the existence of a NT equilibrium has created its own share of confusion in relation with the definition of the entropy \cite{dunkel14,calabrese19}, NTs are now broadly accepted in line with different experimental observations \cite{frenkel15,buonsante16,puglisi17,cerino15,abraham17,
baldovin21,onorato21,comment}. 
NTs were originally observed experimentally in nuclear spin systems \cite{purcell51}.
More recently, NTs were observed with cold atoms in optical lattices \cite{braun13}.
Furthermore, NTs originally predicted by Onsager in the statistical description of point vortices \cite{onsager49} have been recently observed in 2D quantum superfluids \cite{gauthier19,johnstone19}. 

In this Letter we present an experimental optical setup in which we report the observation of light thermalization to NT equilibrium states. 
Our 
system is based on the nonlinear 
propagation of speckle beams in a multimode optical fiber (MMF).
Because of the presence of a finite number of modes supported by the MMF, the spectrum exhibits  both lower and upper bounds for the energy levels.
The bounded spectrum, combined to the nonlinear four-wave interaction, are responsible for the process of Rayleigh-Jeans thermalization to NT equilibrium states \cite{christodoulides19,EPL21}.
We stress that, at variance with other experiments where photon thermalization 
is driven by a thermal heat bath \cite{conti08,weitz,fischer,bloch21}, here light thermalization takes place in a conservative Hamiltonian system.
RJ thermalization to usual {\it positive temperature} equilibriums has been recently demonstrated experimentally in MMFs \cite{PRL20,wise22,pod22,mangini22}, on the basis of a spatial beam-cleaning effect \cite{liu16,wright16,krupa17,pod19,kottos20}.
As described by the wave turbulence theory \cite{zakharov92,nazarenko11,Newell_Rumpf,laurie12,PR14} applied to MMFs \cite{PRA11b,PRL19,PRA19,PRL22}, the thermalization to a positive temperature equilibrium is characterized by a transfer of power (particle number) 
toward the low-order modes of the MMF.
In marked contrast, here we report the observation of thermalization to a NT equilibrium featured by a power transfer to high-order modes  (direct flow of particles), as well as a transfer of energy to low-order modes (inverse flow of energy).
Consequently, the NT equilibrium is characterized by an inverted modal population, in which high-order modes are more populated than low-order modes.

Our experimental optical setup can be used as a simple and flexible testbed to explore fundamental issues related to NT states that are discussed in conclusion, e.g., Carnot cycles operating between temperatures of opposite signs,  
or inverted turbulence cascades featured by an analogue process of condensation at NT.

{\it Experimental system.-} 
The experiment is based on the single pass propagation of speckle beams through a MMF.
The subnanosecond pulses delivered by a Nd:YAG laser ($\lambda=1.06\mu$m) are transmitted through a spiral phase plate and then through a diffuser before injection of the speckle beam into a 12m long graded-index MMF (i.e., parabolic-shaped trapping potential), which guides $M = 45$ modes, i.e., nine groups of degenerate modes.
The energy levels (fiber eigenvalues) are well approximated by the ones of an harmonic potential $\beta_p=\beta_0(p_x+p_y+1)$, where $\{p\}$ labels the two integers $(p_x,p_y)$ that specify a mode (see Supplementary Material).
We denote by $|a_p|^2$ the 
power in the mode $p$, with 
the total power 
$N=\sum_p |a_p|^2$ \cite{agrawal}.

The experiment is realized in the weakly nonlinear regime, where linear effects dominate over nonlinear effects $L_{lin} \sim \beta_0^{-1} \sim 0.1{\rm mm} \ll L_{nl}=1/(\gamma N)\sim 20 {\rm cm}$, $\gamma$ being the nonlinear coefficient of the MMF.
Accordingly, we do not consider NT states associated to 
nonlinear coherent structures, e.g., breathers \cite{iubini13,rumpf09,baldovin21}.
Since $L_{lin} \ll L_{nl}$, we only retain the linear contribution to the Hamiltonian, $E=\sum_p \beta_p |a_p|^2$ \cite{PRL20,wise22,mangini22}.
We have verified the conservation of the power $N$ and the energy $E$ through propagation in the NT region for each realization of a speckle beam, which confirms that the coupling between guided modes and leaky modes of the fiber can be neglected (see Supplementary Material).

RJ thermalization is driven by the four-wave nonlinear interaction 
through the propagation in the MMF. 
The speckle beam is expected to relax toward the thermodynamic equilibrium state described by the RJ distribution \cite{christodoulides19,PRL19,PRL20,PRA19,wise22,mangini22}:
\begin{equation}
n_p^{\rm RJ}=T/(\beta_p-\mu),
\label{eq:rj}
\end{equation}
where $T$ and $\mu$ are the temperature and chemical potential, while $n_p=\left< |a_p|^2 \right>$ denotes the modal power averaged over the realizations of the speckle beams. 
We have at equilibrium $N=T\sum_p (\beta_p-\mu)^{-1}$ and $E=T \sum_p \beta_p /(\beta_p-\mu)$, with $(T,\mu)$ uniquely determined by $(N, E)$ 
-- we deal with a microcanonic description ($T$ is not defined by a thermostat, it is in units of W$\cdot$m$^{-1}$) \cite{PR14}.
Note that the RJ distribution refers to the classical, low-energy, limit of the Bose-Einstein distribution \cite{zakharov92}, describing highly occupied fiber modes.


\begin{figure}
\includegraphics[width=1\columnwidth]{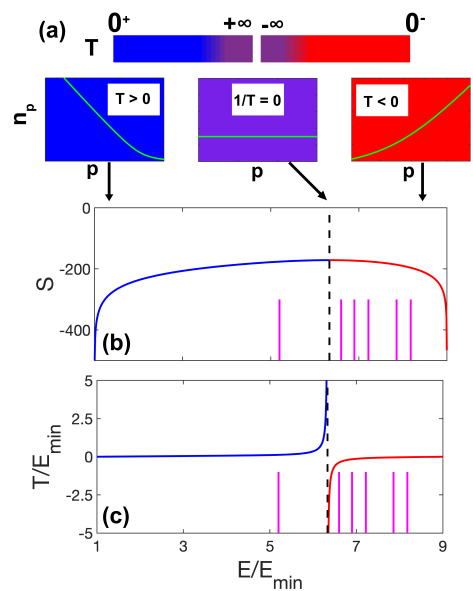}
\caption{
\baselineskip 10pt
{\bf Negative temperatures and inverted modal population.} 
(a) RJ equilibrium distribution $n_p^{\rm RJ}$ for positive temperature $T>0$ ($E<E_*$) where low-order modes are more populated, and negative temperatures $T<0$ ($E>E_*$) featured by an inverted modal population; while for $1/T \to 0$ ($E=E_*$), $n_p^{\rm RJ}=$const.
(b) Relative entropy $S$ vs energy $E$, showing that $1/T=(\partial S/\partial E)_{N,M} < 0$ requires $E>E_*$.
(c) Temperature $T$ vs energy $E$. 
Negative temperatures $T<0$ occur for $E>E_*$ with $E_*/E_{\rm min}\simeq 6.33$ (vertical dashed black line).
The vertical purple lines in (b-c) denote the six values of $E$ considered in Fig.~\ref{fig:np}.
}
\label{fig:NT} 
\end{figure}

{\it Negative temperatures.-} 
The irreversible process of RJ thermalization is described by the wave turbulence theory \cite{zakharov92,nazarenko11,Newell_Rumpf,laurie12,PR14}, which provides a nonequilibrium description of light propagation in MMFs \cite{PRA11b,PRL19,PRA19,PRL22}.
An equilibrium thermodynamic formulation of multimode optical systems has been recently developed \cite{christodoulides19,efremidis21}.
We report in Fig.~\ref{fig:NT} the relative entropy $S=\sum_p \log(n_p^{\rm RJ})$ as a function of the energy for the MMF used in our experiments with $g=9$ groups of degenerate modes.
Because the spectrum of the fiber is bounded, $\beta_0 \le \beta_p \le \beta_{\rm max}=g \beta_0$, the system possesses both lower and upper energy bounds
\begin{equation}
E_{\rm min}=N \beta_0 \le E \le E_{\rm max}=N \beta_{\rm max}. 
\label{eq:boundE}
\end{equation}
Starting at minimum energy $E_{\rm min}$, where only the fundamental mode is populated, an increase in energy leads to an occupation of a larger number of fiber modes and therefore an increase in entropy. 
As the temperature approaches infinity, all fiber modes become equally populated $n_p^{\rm RJ}=$const, and the entropy reaches a maximum 
for $E=E_*=N \left< \beta_p\right>=E_{\rm min}(2 g+1)/3$. 
NT equilibrium states arise for $E > E_*$, where the entropy decreases by increasing the energy, $1/T=(\partial S/\partial E)_{M,N} < 0$.
The condition $E > E_*$ can be achieved if high-order modes are more populated than low-order modes. 
Note that NT equilibrium states persist in the thermodynamic limit (see Supplementary Material).

{\it RJ thermalization to NT equilibriums.-}
At variance with usual experiments of spatial beam cleaning and RJ thermalization \cite{liu16,wright16,krupa17,pod19,wise22,pod22,mangini22}, here we study the thermalization for different values of the energy $E$, while keeping constant the power $N$.
Indeed, by passing the laser beam through a diffuser before injection into the fiber, we can vary the amount of randomness of the speckle beam by keeping $N=$const -- the larger the randomness of the speckle beam, the higher the energy $E$.
Accordingly, we study RJ thermalization over a broad range of variation of the energy.
In order to further increase the energy beyond the threshold for NT ($E>E_*$), we pass the beam through a spiral phase plate before the diffuser, i.e., we generate a speckle beam from a doughnut-like intensity distribution, which enables the excitation of higher order fiber modes. 

\begin{figure}
\includegraphics[width=1\columnwidth]{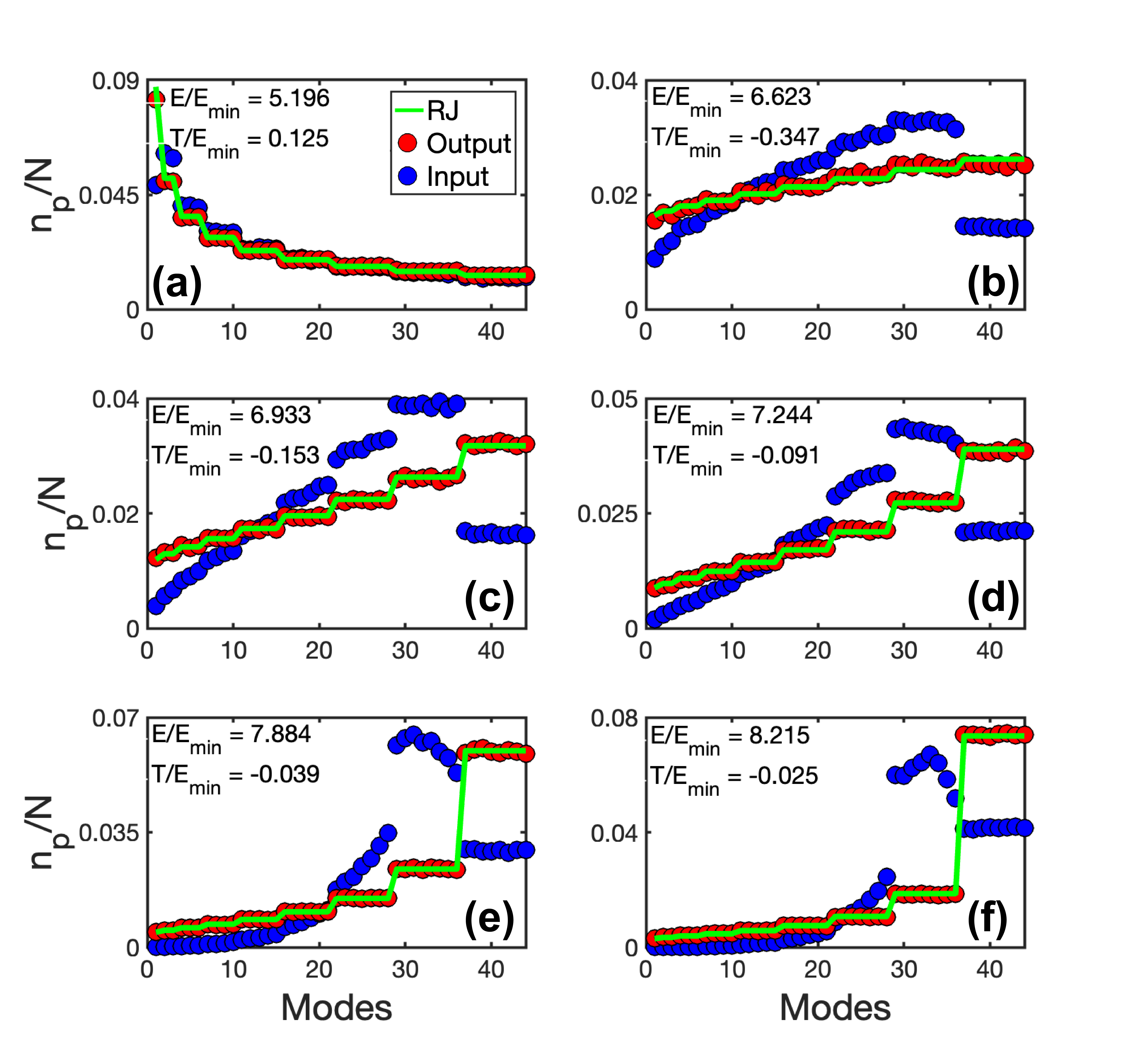}
\caption{
\baselineskip 10pt
{\bf Rayleigh-Jeans thermalization to NT equilibriums.} 
Experimental modal distributions averaged over realizations $n_p^{\rm exp}=\left<|a_p^{\rm exp}|^2\right>$, 
at the fiber input (blue), 
at the fiber output (red).
Corresponding RJ equilibrium distribution $n_p^{\rm RJ}$ (green).
Note the quantitative agreement between $n_p^{\rm RJ}$ and the experimental output distribution $n_p^{\rm exp}$ (red).
The six panels correspond to different values of $E$, or equivalently different $T$,
see the six vertical purple lines in Fig.~\ref{fig:NT}(b-c).
The modal distribution peaked on the lowest mode for $T>0$ (a), gets inverted for $T<0$ (b-f).
An average over $\simeq$35 realizations of speckle beams is considered for each panel.
The fiber modes are sorted from the fundamental one ($\beta_0$) to the highest mode group (nine-fold degenerate with $\beta_{\rm max}=9\beta_0$).
Degenerate modes are equally populated at equilibrium, leading to a staircase distribution $n_p^{\rm RJ}$.}
\label{fig:np} 
\end{figure}

The accurate measurements of the near-field and far-field intensity distributions allowed us to retrieve the modal power distribution $n_p^{\rm exp}$.
To obtain the mode decomposition, several interferometric approaches based on use of a reference beam have been exploited to study light thermalization in MMFs \cite{wise22,mangini22,pod22}.
Here, in contrast to the previous works \cite{PRL20,wise22,mangini22,pod22}, we use a non-interferometric numerical mode decomposition procedure that is based on the 
Gerchberg-Saxton algorithm.
It allows us to retrieve the transverse phase profile of the speckle field from the near-field and far-field  intensity distributions measured in the experiments \cite{fienup82,shapira05,shechtman15,turitsynNC20}.
By projecting the retrieved complex field 
over the fiber modes, we get the complete modal distribution.

\begin{figure}
\includegraphics[width=1\columnwidth]{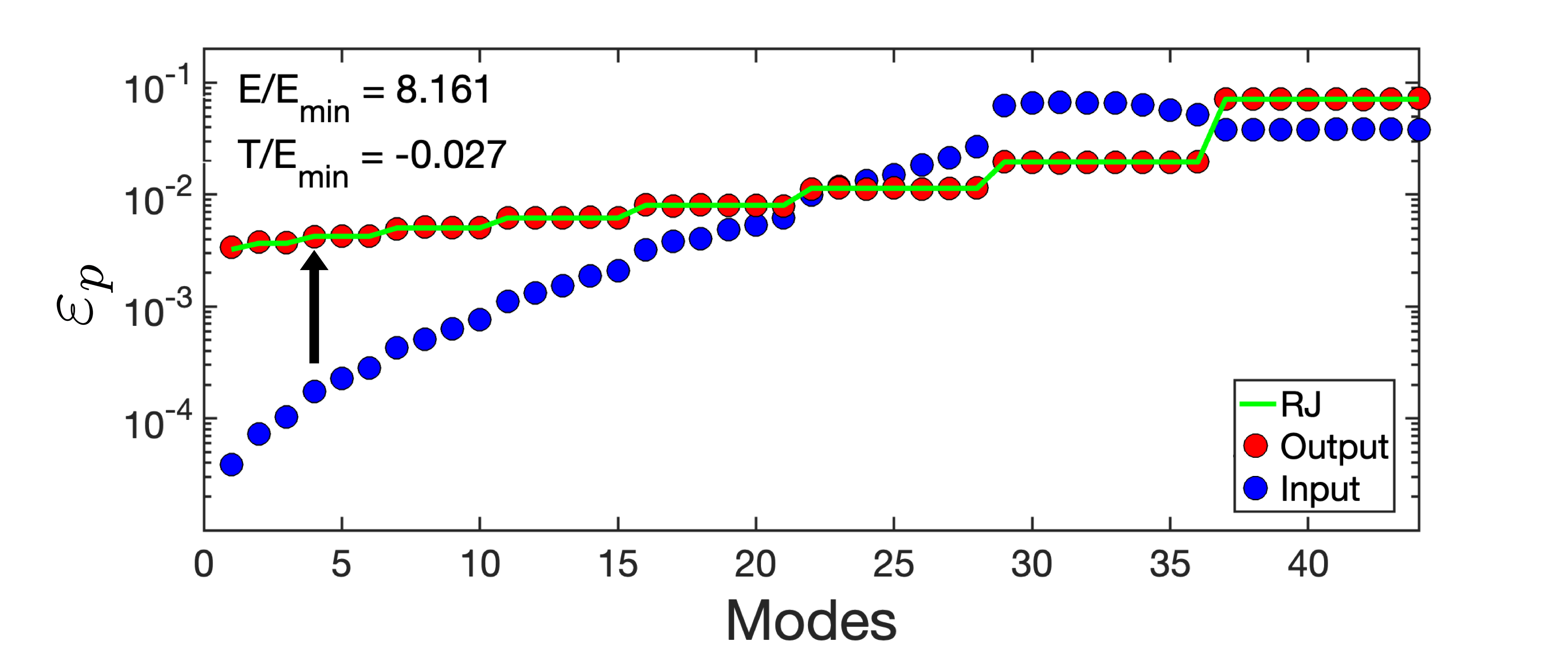}
\caption{
\baselineskip 10pt
{\bf Energy flows in mode space.} 
Experimental energy distributions averaged over $50$ realizations $\varepsilon_p^{\rm exp} =\beta_p n_p^{\rm exp}$, at the fiber input (blue), 
output (red).
The arrow indicates the energy flow to low-order modes.
Corresponding RJ equilibrium distribution $\varepsilon_p^{\rm RJ}=\beta_p n_p^{\rm RJ}$ (green line), which is in quantitative agreement with the experimental output distribution (red).
}
\label{fig:E} 
\end{figure}

The RJ distribution being in essence a statistical distribution, its comparison with the experiments requires an average over realizations of speckle beams.
We have recorded 2$\times$300 realizations of the near-field and far-field intensity distributions for the same power ($N=7$kW) and different energies $E$.
For each individual speckle realization, we  retrieve the modal distribution $|a_p^{\rm exp}|^2$.
We partition the ensemble of 300 realizations of $\{|a_p^{\rm exp}|^2\}$ within small energy intervals $[E-\Delta E, E+\Delta E]$ with $\Delta E=0.125 E_{\rm min}$.
We  perform an average over the realizations of the modal distributions for each energy interval, which provides the averaged modal distribution $n_p^{\rm exp}=\left<|a_p^{\rm exp}|^2\right>$.
This procedure is applied at the fiber output ($L=12$m), and fiber input (after 20cm of propagation).
The error in the procedure has been computed theoretically and numerically, 
it decreases with the number of realizations and has been found remarkably small (relative standard deviations of $\simeq 6\%$), see Supplementary Material.

We report in Fig.~\ref{fig:np} the averaged modal distributions $n_p^{\rm exp}$ at the fiber input (blue) and output (red), for different values of the energy $E$, or equivalently  the temperature $T$ (purple lines in Fig.~\ref{fig:NT}(b-c)).
The 
data are compared with the theoretical RJ distribution $n_p^{\rm RJ}$.
We stress that there are no adjustable parameters between 
$n_p^{\rm exp}$ and 
$n_p^{\rm RJ}$: 
The parameters $(T,\mu)$ in $n_p^{\rm RJ}$ 
are uniquely determined by 
$N$ and $E$ measured in the experiments.
We observe in Fig.~\ref{fig:np} an excellent agreement between ${n_p^{\rm exp}}$ (red circles) and $n_p^{\rm RJ}$ (green line), for both $T>0$ and $T<0$.
Fig.~\ref{fig:np} then shows that NT equilibriums constitute {\it attractor states} for the random wave, whose robustness has a thermodynamic origin -- maximum entropy state for a given pair $(N,E)$.

{\it Energy flows in mode space.-}
The conventional thermalization to positive temperatures is 
characterized by an energy flow to high-order modes \cite{nazarenko11,Newell_Rumpf,PRL19}.
Thermalization to NTs typically occurs through an inverse energy flow to low-order modes \cite{EPL21}. 
This is illustrated in Fig.~\ref{fig:E}, which shows that the energy distribution $\varepsilon_p^{\rm exp} = \beta_p n_p^{\rm exp}$ at low-order modes increases through propagation in the MMF and reaches the theoretical RJ equilibrium distribution $\varepsilon_p^{\rm RJ} = \beta_p n_p^{\rm RJ}$.

\begin{figure}
\includegraphics[width=1\columnwidth]{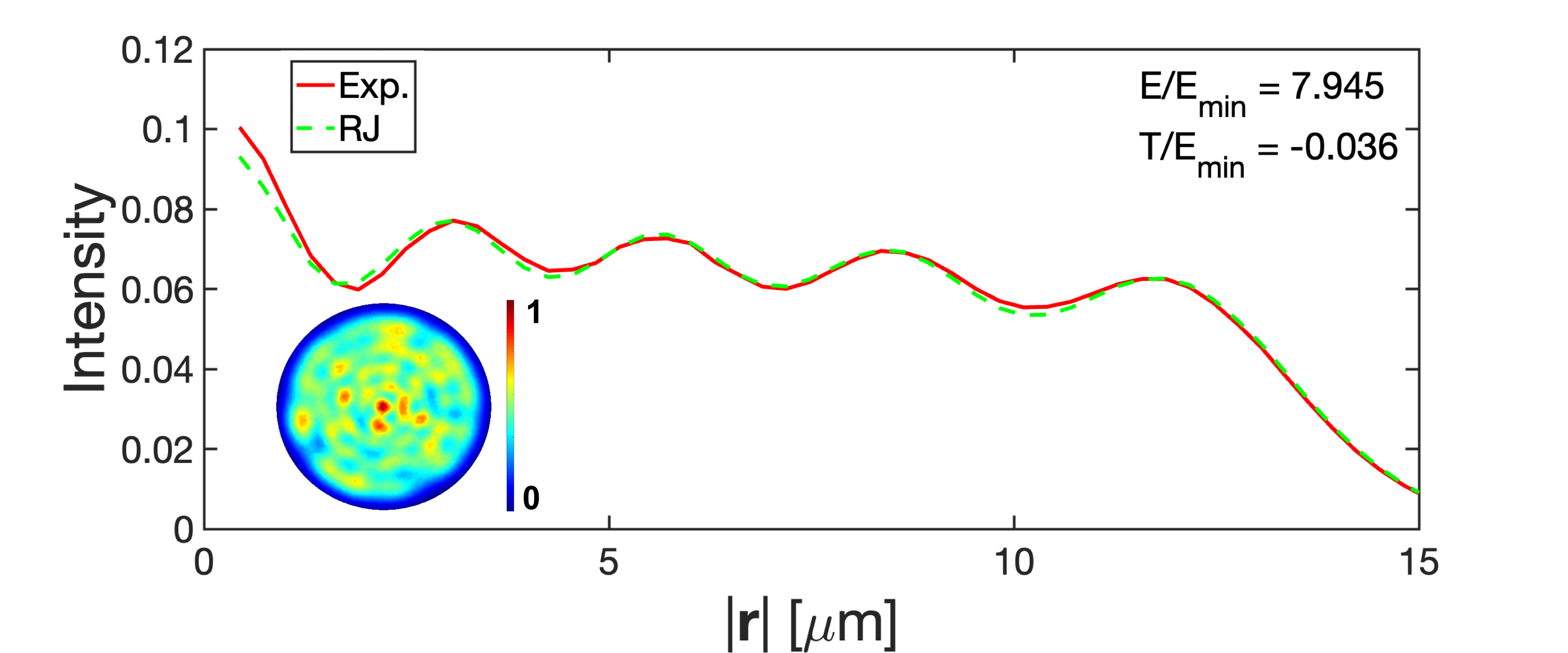}
\caption{
\baselineskip 10pt
{\bf Oscillating radial intensity distribution at NT.} 
Averaged intensity distribution $I^{\rm exp}(|{\bm r}|)$ as a function of the radial (angle-averaged) distance $|\br|$ (red).
Note the quantitative agreement with the theoretical RJ intensity distribution $I^{\rm RJ}(|\br|)$ in Eq.(\ref{eq:int_rj}) (dashed green).
The oscillating behavior of the intensity distribution is a signature of the NT equilibrium.
Inset: corresponding 2D intensity averaged over the realizations (the radius of the circle is the fiber radius).
}
\label{fig:I} 
\end{figure}

\newpage
{\it Oscillating radial intensity distribution.-}
The intensity distribution $I^{\rm RJ}(|{\bm r}|)$ of usual positive temperature equilibriums is, in average, a monotonic decreasing function with the radial distance $|{\bm r}|$ \cite{PRL20}.
This is consistent with the intuitive idea that low-order modes localized near-by the fiber center are the most populated ones.
In marked contrast, the inverted modal population of NT equilibriums are characterized by an oscillating behaviour of the radial intensity distribution.
This is illustrated in Fig.~\ref{fig:I}, which reports the averaged radial intensity distribution $I^{\rm exp}(|{\bm r}|)$ 
(with $\Delta E=0.25 E_{\rm min}$, $E/E_{\rm min}=7.9$).
The theoretical RJ intensity distribution reads
\begin{equation}
I^{\rm RJ}(\br)=\sum_{p} n_{p}^{\rm RJ} u_{p}^2(\br), 
\label{eq:int_rj}
\end{equation}
where $u_p(\br)$ denotes the 
fiber modes \cite{PRL20}.
The number of radial oscillations in Fig.~\ref{fig:I} is given by the most oscillating mode of the fiber, namely the mode LP$_{04}$ that exhibits 5 oscillations.

\begin{figure}
\includegraphics[width=1\columnwidth]{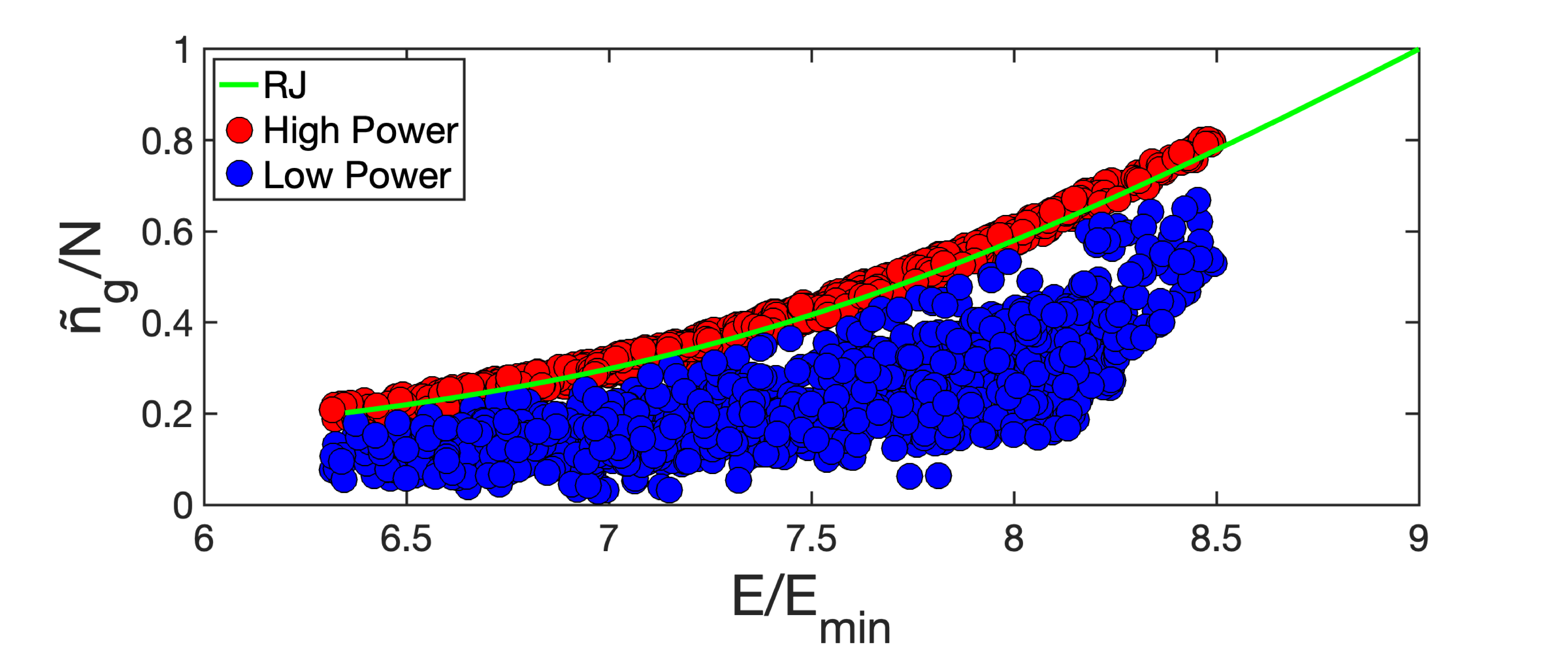}
\caption{
\baselineskip 10pt
{\bf Macroscopic population of the highest energy level.} 
Fraction of power ${\tilde n}_g/N$ 
into the highest mode group $g=9$ vs energy $E/E_{\rm min}$.
Experimental measurements at the fiber output: The blue circles refer to the linear regime (small power), the red circles to the nonlinear regime (high power).
The green line denotes the RJ equilibrium theory.
By increasing the energy, the power goes to the highest energy level, ${\tilde n}_g/N \to 1$ as $E/E_{\rm min} \to 9$.
}
\label{fig:P} 
\end{figure}

{\it Experiments by increasing power.-}
We have studied the optical field at the output of the MMF, with a small power $N=0.23$kW (linear regime), and a high power $N=7$kW (nonlinear regime).
Since the MMF length is kept fixed ($L=12$m), the effective number of nonlinear interaction lengths increases by increasing the power. 
Fig.~\ref{fig:P} reports the fraction of power that populates the highest group of degenerate modes of the MMF,  
${\tilde n}_g/N$ for $g=9$. 
The output field (red) reaches the equilibrium RJ theory (green line) in the nonlinear regime. 
The highest energy level gets macroscopically populated by increasing the energy, or equivalently by increasing the temperature of negative sign (see Fig.~\ref{fig:NT}).

{\it Conclusion and perspectives.-}
We have reported the observation of RJ thermalization to NT equilibrium states through light propagation in graded-index MMFs. 
This non-equilibrium process of NT thermalization can be described by a wave turbulence kinetic equation, which is found in agreement with the simulations of the nonlinear Schr\"odinger equation (see Supplementary Material). 
Our NT experiment then paves the way for the study of Zakharov-Kolmogorov turbulence cascades \cite{zakharov92,nazarenko11,Newell_Rumpf} that are inverted with respect to those underlying usual positive temperature thermalization (e.g., inverse energy flow in Fig.~\ref{fig:E}).

Along this line, our work suggests a previously unrecognized process of inverted condensation at NTs: At variance with usual condensation at positive temperature where the lowest energy level gets macroscopically populated by {\it decreasing} the temperature ($T \to 0^+$, or $E \to E_{\rm min}$) \cite{nazarenko11,Newell_Rumpf,laurie12,PR14,PRL20}, at NT an inverted condensation process occurs into the highest energy level as the temperature {\it increases to zero} ($T \to 0^-$, or $E \to E_{\rm max}$). 
While we provide a preliminary study of this effect through the macroscopic population of the highest energy level (Fig.~5), the observation of the transition to condensation requires MMFs with larger number of modes (see Supplementary Material).

In our work NT states are obtained directly, which is in contrast with magnetic systems and cold atoms 
where the excitation of NT states requires first the creation of a positive temperature state and then its subsequent inversion through suitable procedures (magnetic field inversion or Feshbach resonances).
This opens the possibility to study the physics of NT in a flexible experimental environment.
For instance, the thermalization of two beams at different laser wavelengths interacting through the fiber nonlinearity can be exploited to achieve an efficient optical 
refrigeration: 
A highly incoherent speckled beam at NT can be cooled through its thermalization with a coherent beam towards a highly coherent state without any power loss.
In contrast with usual beam cleaning at positive temperature where the energy is conserved, here the cooling process is featured by an energy transfer from the incoherent to the coherent beam, which significantly improves the gain of coherence of the incoherent beam.
Following this idea, one can explore the meaning of a {\it thermostat at NT} \cite{baldovin21}: If the NT incoherent beam has a power much larger than the partially coherent beam, it will play the role of a NT thermal reservoir for such a partially coherent beam.

The versatile optical experimental environment proposed in this work also opens the possibility to study controversies about NTs, such as thermodynamic engines featured by Carnot cycles operating between temperatures of opposite signs, in relation with the generalized Kelvin-Planck formulation of the second law of thermodynamics stating that it is not possible to completely transform work into heat at NT \cite{baldovin21}.

{\it Acknowledgments.-} The authors are grateful to S. Rica, I. Carusotto and V. Doya for fruitful discussions. Fundings: Centre national de la recherche scientifique (CNRS), Conseil r\'egional de Bourgogne Franche-Comt\'e, iXCore Research Fondation, Agence Nationale de la Recherche (ANR-19-CE46-0007, ANR-15-IDEX-0003, ANR-21-ESRE-0040). Calculations were performed using HPC resources from DNUM CCUB (Centre de Calcul, Universit\'e de Bourgogne).

\section{Supplementary Material}

\section{Experimental set-up}
\label{sec:setup}

\begin{figure}
\includegraphics[width=.7\columnwidth]{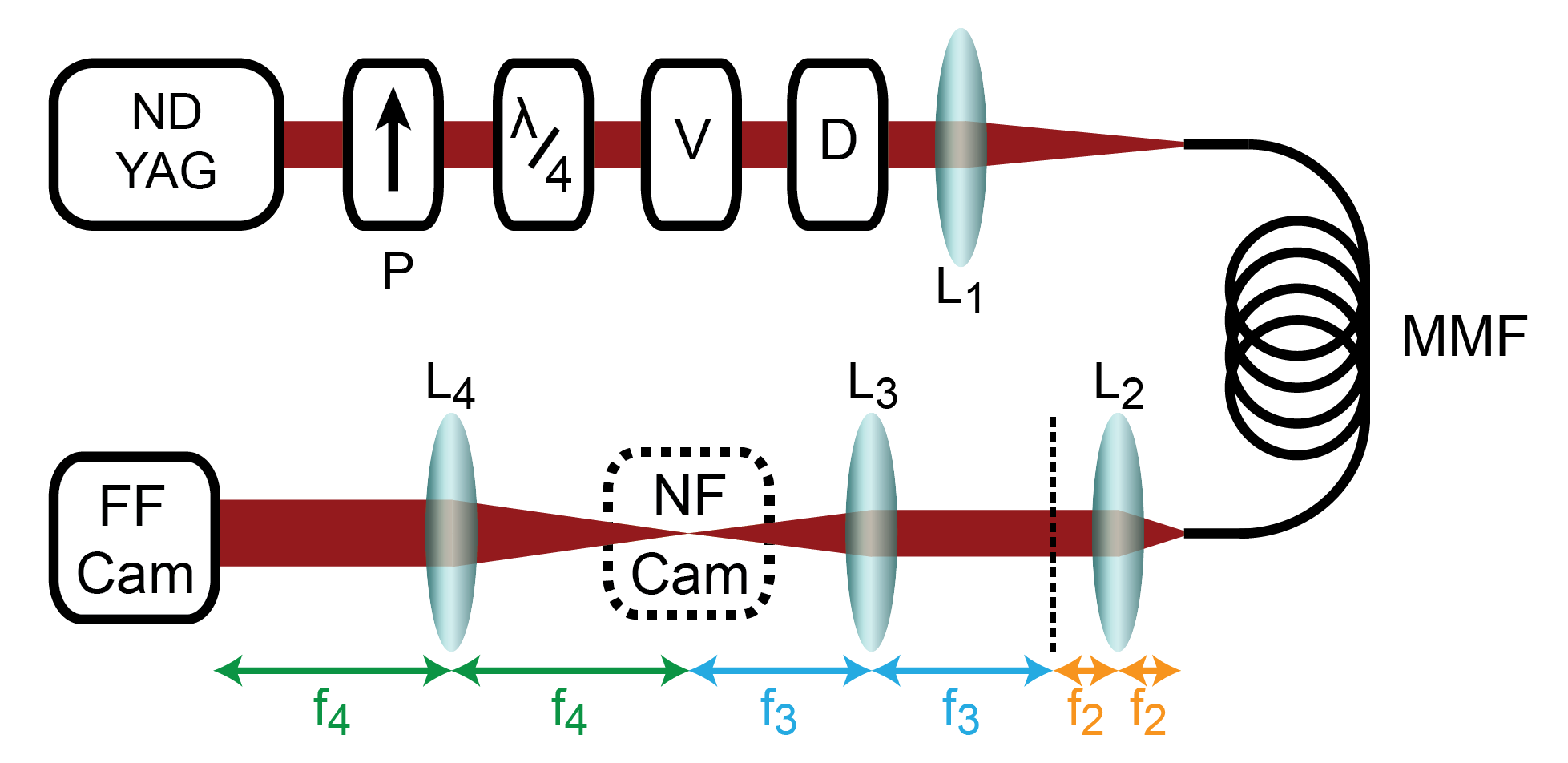}
\caption{
\baselineskip 10pt
{\bf Setup.} 
laser, optical isolator,
half-wave plate and polarizer, lenses for magnification and
imaging ($f_j$), spiral phase plate (V), diffuser (D), graded-index MMF, and cameras for near- and far-field detections (Cam).
}
\label{fig:setup} 
\end{figure}

The source is a Nd:YAG laser delivering subnanosecond pulses (400ps) at $\lambda=$1064nm. 
The laser beam is passed through a spiral phase plate (Thorlabs) to generate a doughnut-like ring-shaped beam, and subsequently through a diffuser before injection of the speckle beam into the MMF, see Fig.~\ref{fig:setup}.
The diffuser plate is placed in the vicinity of the Fourier plane of a 4f-optical system. 
The near-field (NF) intensity distribution of the fiber output beam was magnified and imaged on a first CCD camera owing to a two lens telescope optical system, with $f_2$ = 8 mm and $f_3$ = 150 mm. 
The CCD camera was placed on a rail orthogonal to the beam propagation in order to remove or put the camera back on the beam path. 
The far-field (FF) intensity distribution of the magnified image was obtained by placing it in the object focal-plan of a lens 
$f_4$ = 250 mm and using a second CCD camera positioned in its image (Fourier) focal-plan.

We have computed analytically the propagation of the optical wave throughout the setup of our detection scheme, according to Fig.~\ref{fig:setup} (lower part). 
If $\psi_0(\br)$ is the optical field amplitude at the fiber output ($\br=(x,y)$), then we have in the NF plane:
$$
\psi_{\rm NF}(\br) = - \rho^{-1} \psi_0(-\br/\rho) ,
$$
with $\rho=f_3/f_2$ the magnification factor.
In the FF plane, the wave amplitude reads
$$
\psi_{\rm FF}({\bm u}) = \frac{i \rho}{\lambda f_4}  \int d\br 
\psi_0(\br) \exp[- i 2 \pi (-\rho) \br \cdot {\bm u} / (\lambda f_4) ],
$$ 
which corresponds to the Fourier transform of the field amplitude at the fiber output (note that the constant phase prefactor plays no role because the camera records the intensity).
We note that: (i) The optical amplitude in the NF plane is an exact magnification of the wave amplitude at the output of the MMF; (ii) the optical amplitude in the FF detection plane exactly corresponds to the Fourier transform of the amplitude at the fiber output.
Then, the experimental setup for the detection of the NF and FF intensities does not introduce detrimental spurious transverse phases profiles in the optical field, e.g., related to optical free propagation in air or phase shifts due to the presence of additional lenses.

\baselineskip 11pt

\noindent \smallskip 

\subsection{Multimode fiber}
 
The refractive index profile of the graded-index MMF exhibits a parabolic shape in the fiber core with a maximum core index (at the center) of $n_{\rm co}\simeq$1.472 and $n_{\rm cl} \simeq 1.457$ for the cladding at the pump wavelength of 1064nm (fiber radius $R=15 \mu$m).
The fiber length is $L=12$m.
The trapping parabolic potential reads $V(\br)=q |\br|^2$ for $|\br| \le R$ and 
$q=k_0(n_{\rm co}^2-n_{\rm cl}^2)/(2 n_{\rm co} R^2)$, $k_0=2\pi/\lambda$ the laser wave-number.
The fundamental mode energy level is $\beta_0=2\sqrt{\alpha q}$, with $\alpha=1/(2 n_{\rm co} k_0)$.
The MMF guides $M = 45$ modes (i.e. $g=9$ groups of degenerate modes).

\begin{figure}
\includegraphics[width=1\columnwidth]{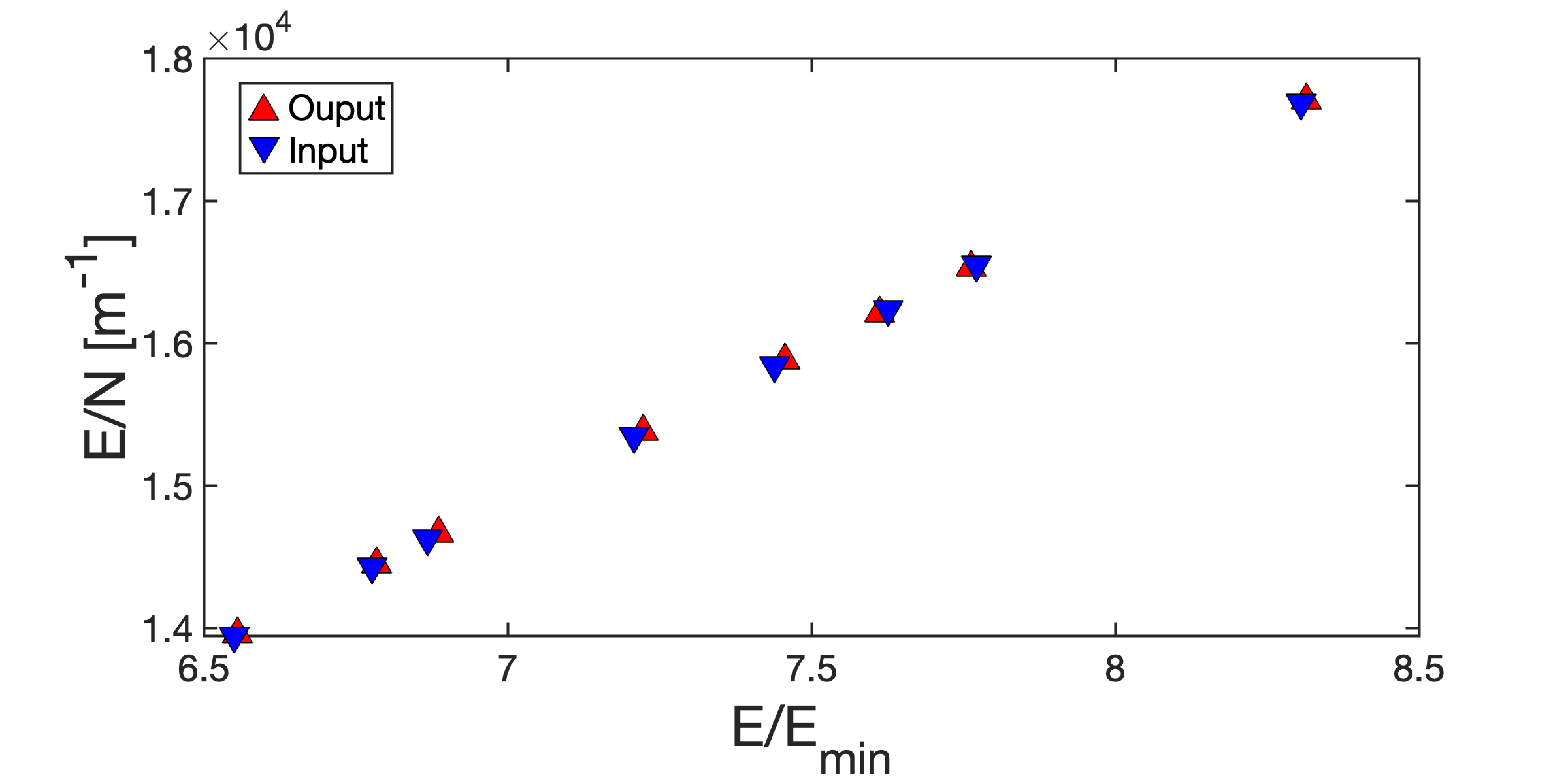}
\caption{
\baselineskip 10pt
{\bf Conservation of the energy through propagation in the MMF in the negative temperature region.} 
(a) Measurements of the energy at the input of the MMF  (blue triangles), and at the output of the MMF (red triangles): The energy $E/N$ is conserved through the propagation in the MMF over a broad range of variation of $E/E_{\rm min}$.
We recall that negative temperatures $T<0$ occur for $E>E_*$ with $E_*/E_{\rm min}\simeq 6.33$, see Fig.~1.
}
\label{fig:exp_E} 
\end{figure}

The truncation of the potential introduces a frequency cut-off in the FF spectrum $k_c=(2\pi/\lambda)\sqrt{n_{\rm co}^2-n_{\rm cl}^2}$ \cite{PRL20}.
The conservation of $N$ and $E$ through propagation in the MMF (see Fig.~\ref{fig:exp_E}) shows that the coupling from guided modes to leaky radiation modes in the cladding is negligible.
This is not surprising as the efficiency of such a coupling can be shown to be very small.
Indeed, the MMF has a core (radius 15$\mu$m), a cladding (radius 62.5$\mu$m), 
and a highly absorbing polymer-coating with refractive index {\it larger than the core}.
Then leaky radiation modes in the cladding are rapidly absorbed during propagation due to their large penetration in the polymer-coating: 
we measured a typical absorption length $L_{abs}$ of $\simeq 15$cm. 
Let us consider the coupled amplitude equations describing the four-wave mixing between four modes including a leaky mode. The four mode amplitudes $a_1,a_2,a_3,a_{4}$ (where $a_{4}$ stands for the leaky mode amplitude) satisfy Eqs.(10.2) in \cite{agrawal}, in which we need to add an absorption term of the form $-a_{4}/L_{abs}$ in the equation (10.2.4) in \cite{agrawal}  for $a_{4}$.
Since absorption is strong the overdamped limit is valid and we get 
$a_{4}= i n_2 k_0 L_{abs} f_{4312} a_1 a_2 a_3^* \exp (-i \Delta \beta z)$ where $\Delta \beta= \beta_3+\beta_4-\beta_1-\beta_2$, the parameter $n_2$ is the nonlinear-index coefficient, and $f_{4312} $ is an overlap integral between the four mode profiles.
By substitution into one of the first three equations, say (10.2.1) in \cite{agrawal}, we find that the mode amplitude $a_1$ experiences an effective absorption due to the coupling with the leaky mode that is given by $-4 n_2^2 k_0^2 L_{abs} |f_{1234}|^2 |a_2|^2 |a_3|^2 a_1$. This shows that this absorption is of the order of $-L_{abs}/L_{nl}^2 a_1$ times a coefficient that is of the order of the square overlap integral between guided and leaky mode profiles.
As the supports of these modes are very different (the guided modes are essentially supported in the core while the leaky modes are essentially supported in the cladding that is much larger), the square overlap integrals are small (smaller than the respective core-cladding  ratio $(15/62.5)^4 \simeq 3 \, 10^{-3}$) and the {\it effect of the coupling to the leaky modes onto the guided mode amplitudes can be neglected} when $L_{nl}\simeq 20$cm and the propagation distance is $L=12$m.

\noindent \bigskip 

\subsection{Measurements of the energy $E$}

From the measurements of the NF and FF intensity distributions, we have retrieved an accurate measurement of the power $N$ and the energy $E$ of the speckle beam.
The NF intensity distribution $I_{\rm NF}(\br)=|\psi|^2(\br)$ provides a measurement of the power $N=\int I_{\rm NF}(\br) d\br$ and of the potential energy $E_{\rm pot}=\int V(\br) |\psi|^2(\br) d\br$.
The kinetic energy $E_{\rm kin}=\alpha \int |\nabla \psi|^2(\br) d\br$ is retrieved from the FF intensity distribution $I_{\rm FF}(\bk)=|{\tilde \psi}|^2(\bk)$.
This provides the measurement of the (linear) energy (Hamiltonian) $E=E_{pot}+E_{kin}$.

\subsection{Conservation of $N$ and $E$ through propagation in the MMF for $E>E_*$ (negative temperature region)}
 
Power conservation has been verified by keeping fixed the conditions of injection of the speckle beam into the MMF: We measured $N_{out}$ at $L=12$m, and then $N_{in}$ by cutting the fiber at $20$cm, and we always obtained $(N_{\rm in}-N_{\rm out})/N_{\rm moy} < 1$\%.
The experimental verification of energy conservation  requires both the NF and FF intensity measurements.
The NF and FF intensities are recorded at the fiber output at $L=12$m, which gives $E_{\rm out}$.
Without altering the fiber launch conditions, the fiber is cut to $20$cm to record the input NF and FF intensities, which gives $E_{\rm in}$.
The measurements of $E_{\rm in}$ and $E_{\rm out}$ then refer to an individual realization of the speckle beam (without average over the realizations).
Fig.~\ref{fig:exp_E} shows that the conservation of the energy is well verified for $E > E_*$, i.e. in the negative temperature region.
The energy $E$ is varied owing to the diffuser before injection into the MMF, see Fig.~\ref{fig:setup}.

\section{Modal decomposition}


\subsection{Phase retrieval}
 
The procedure of mode decomposition is based on the well-known Gerchberg-Saxton algorithm \cite{fienup78,fienup82,shechtman15}.
From the measurements of the NF and FF intensity distributions in the experiment, it allows us to retrieve the transverse phase profile of the field. The resulting complex field is subsequently projected onto the fiber modes, to get the complete modal distribution of the experimental optical beam. The algorithm is known to be accurate although it is not efficient in terms of computational cost. Indeed it is a local search algorithm that updates iteratively the unknown phase profile of the field and it is usually necessary to consider several initial phase guesses. We have, therefore, carried out a detailed preliminary analysis of the algorithm by performing numerical simulations that reproduce our experimental configuration in order to prove that the phase retrieval and modal distribution estimation are reliable.


\begin{figure}
\includegraphics[width=1\columnwidth]{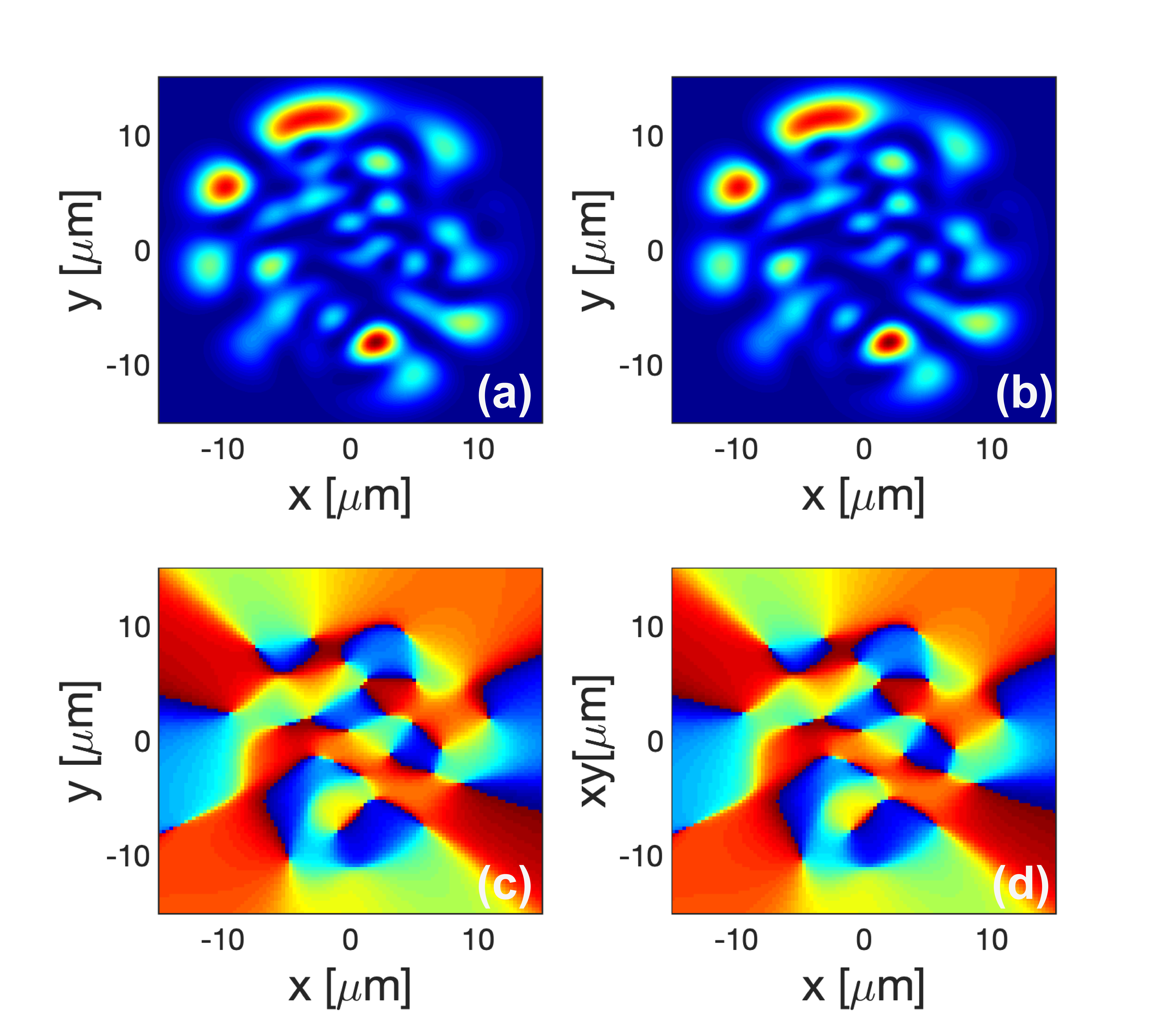}
\caption{
\baselineskip 10pt
{\bf Phase retrieval.} 
Example of numerically generated near-field intensity distribution (a), and its reconstruction (b). Original phase field (c) and the corresponding phase field reconstructed from the Gerchberg-Saxton algorithm (d). 
}
\label{fig:ret_phase} 
\end{figure}


\begin{figure}
\includegraphics[width=.9\columnwidth]{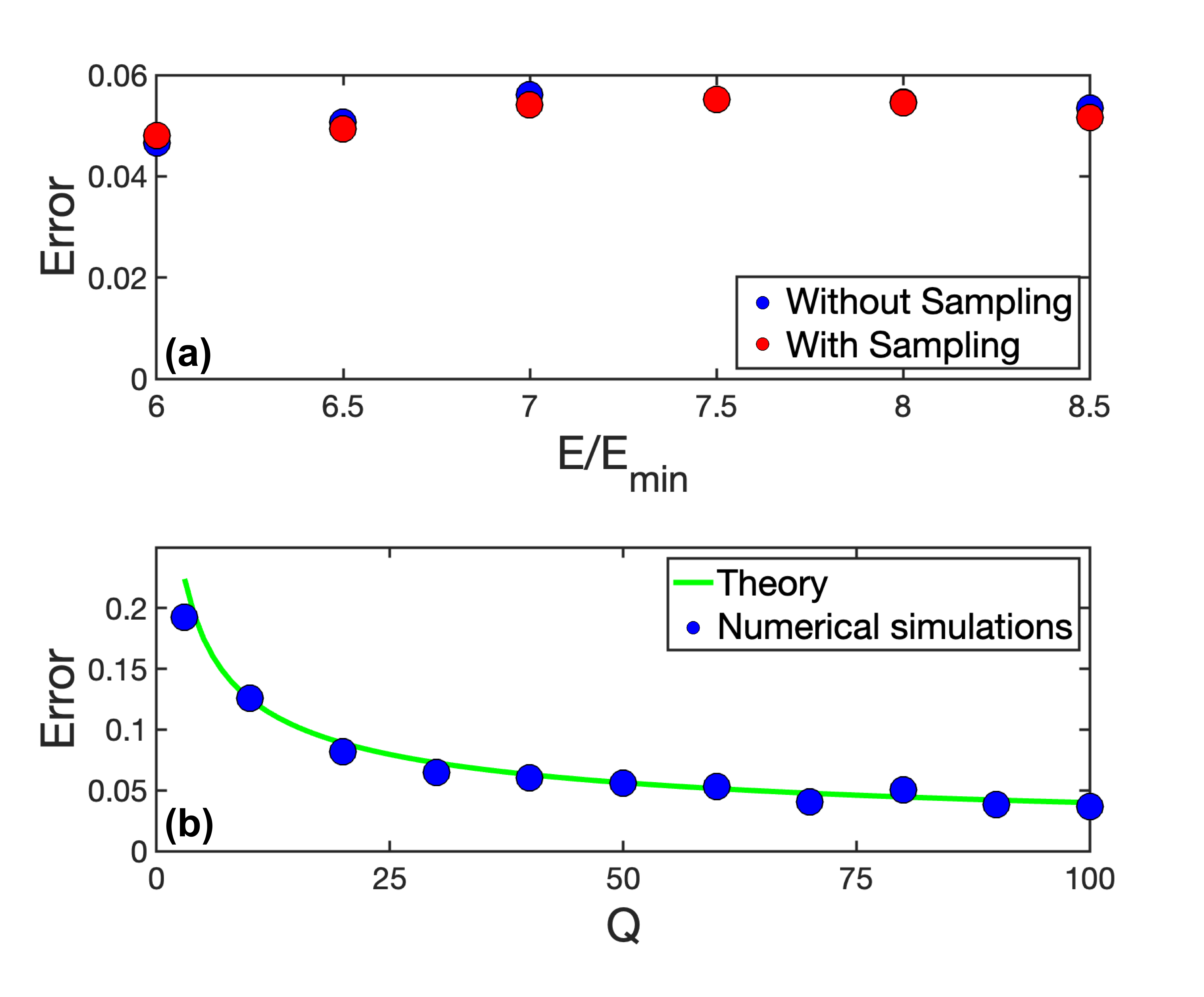}
\caption{
\baselineskip 10pt
{\bf Error in the modal decomposition.} 
The mode decomposition is based on the Gerchberg-Saxton algorithm, whose error 
is quantified by the distance ${\cal D}^Q_{\rm err}$ to the exact distribution, see Eq.(\ref{eq:distance_sampl}).
(a) ${\cal D}^Q_{\rm err}$ vs the energy $E$ (for $Q=50$ realizations), by accounting for the sampling due to the limited resolution of the camera (red), and without the sampling (blue). 
Note in (a) that an increase of the randomness of the speckle beam (i.e., increase of $E$) does not increase the error. 
(b) ${\cal D}^Q_{\rm err}$ vs number of realizations $Q$ (for $E/E_{\rm min}=7$): The error decreases with the number $Q$ of realizations of the speckles.
The green line reports the theoretical estimate of the error given in Eq.(\ref{eq:errQtheo}).
Note that ${\cal D}^Q_{\rm err}$ is bounded, $0 \le {\cal D}_{\rm err}^Q \le 1$.
}
\label{fig:D_vs_E} 
\end{figure}

\begin{figure}
\includegraphics[width=1\columnwidth]{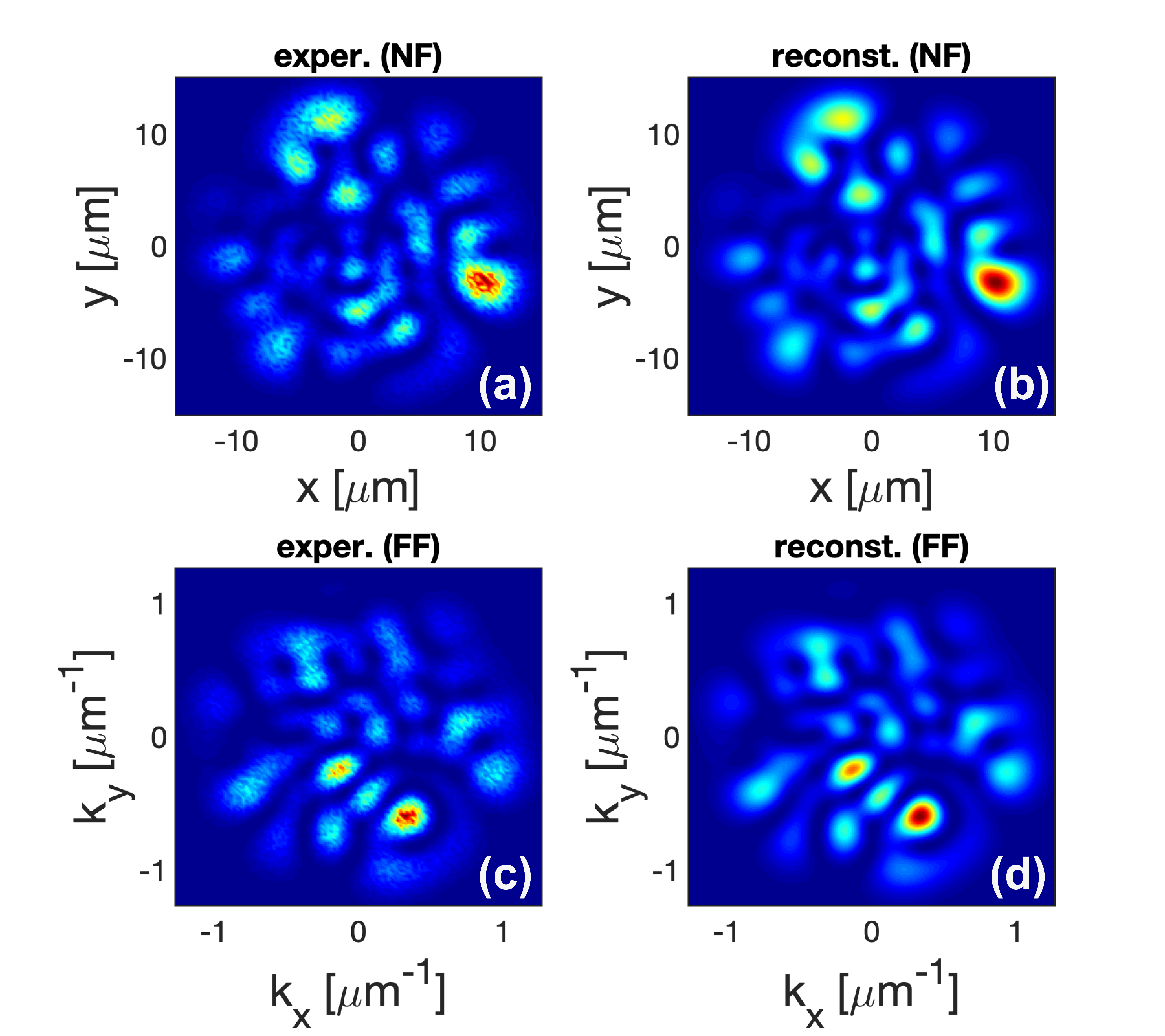}
\caption{
\baselineskip 10pt
{\bf Near-field and far-field experimental intensities, and corresponding reconstructed intensity distributions.} 
Near-field (NF) intensity recorded in the experiment for a single realization of a speckle (a), and corresponding far-field (FF) intensity distribution (c).
Corresponding NF intensity (b), and FF intensity (d), reconstructed from the Gerchberg-Saxton algorithm.
}
\label{fig:2D} 
\end{figure}


\subsection{Error introduced by the Gerchberg-Saxton algorithm}
 
To evaluate accurately the error in the phase-retrieval algorithm, we have reproduced in detail the experimental procedure as follows:

\smallskip 
\noindent
{\bf i)} We consider a particular value of the energy $E$ ($E >E _*$ so that $T < 0$). Throughout the procedure the power $N$ is set constant. The pair $(E,N)$ determines uniquely $(T , \mu)$ and thus the exact RJ distribution at equilibrium $n_p^{\rm RJ}=T/(\beta_p -\mu)$.

\smallskip 
\noindent
{\bf ii)} We generate from $n_p^{\rm RJ}$ a realization of speckle beam $\psi({\br}) =\sum_p a_p u_p(\br)$, where $a_p$ is a complex Gaussian random variable with variance $\left<|a_p|^2\right>=n_p^{\rm RJ}$ 
($a_p=a_p^{(r)}+ia_p^{(i)}$ with $a_p^{(r)}$ and $a_p^{(i)}$ real Gaussian independent random variables with mean zero and variance $n_p^{\rm RJ}/2$).
We recall that $u_p(\br)$ are the fiber modes. Then $\psi(\br)$ is a particular realization of a complex speckle field at exact RJ equilibrium.

\smallskip 
\noindent
{\bf iii)} The particular realization $\psi(\br)$ is highly resolved numerically. We mimic the impact of the finite resolution of
the camera used in the experiment. From $\psi(\br)$  we compute the NF and FF intensity distributions $|\psi(\br)|^2$ and $|\hat{\psi}(\bk)|^2$. We sample the NF and FF intensity distributions with the finite number of
points available in the camera ($1024^2$) in $\br$-space and $\bk$-space
and $\simeq 950$ points for the dynamics range in intensity. We apply the Gerchberg-Saxton algorithm to retrieve the sampled phase profile. 
Due to the errors introduced by the sampling of the camera and by the phase-retrieval algorithm, 
the resulting complex field  $\psi^{\rm exp}(\br)$ may differ from the generated speckle beam $\psi(\br)$.\\
We report in Fig.~\ref{fig:ret_phase} the numerically generated near-field intensity distribution in one numerical simulation (a) and its reconstruction (b), the original phase profile (c) and the reconstructed phase profile (d). It is clear that the reconstruction is very good. We will see below more quantitatively that the error is indeed negligible.

\smallskip 
\noindent
{\bf iv)} We project the complex field $\psi^{\rm exp}(\br)$ onto the fiber modes to get the complex modal coefficient $a_p^{\rm exp}$ and distribution $|a_p^{\rm exp}|^2$.
Due to the errors introduced by the sampling of the camera and by the phase-retrieval algorithm, this modal distribution may differ from the mode distribution $|a_p|^2$ used to generate the speckle beam  $\psi(\br)$. As we will show below, this error is negligible.

\smallskip 
\noindent
{\bf v)} We repeat the steps ii)-iv) $Q$ times, each with a different realization of the speckle beam (i.e., with a different realization $a_p^j$ of $a_p$, for $j=1,\ldots,Q$).
The procedure then gives $Q$ distributions
$|a_p^{{\rm exp},j}|^2$, $j = 1, \ldots, Q$. 
We compute the empirical averages ${n}_p^{{\rm exp},Q} = (1/Q) \sum_{j=1}^Q  |a_p^{{\rm exp},j}|^2$.
We anticipate that, for $Q$ large enough, these empirical averages should be close to the theoretical values $n_p^{\rm RJ}$.
We introduce  the estimation error:
\begin{equation}
{\cal D}_{\rm err}^Q = \frac{ \sum_p \big|{n}_p^{{\rm exp},Q} -
{n}_p^{\rm RJ} \big|}{  \sum_p 
 {n}_p^{{\rm exp},Q} +
{n}_p^{\rm RJ} } .
\label{eq:distance_sampl}
\end{equation}

Let us imagine for a while that the phase-retrieval algorithm is perfect and that the sampling error due to the camera is absent.
Then, for each realization $j=1,\ldots,Q$, we have
$|a_p^{{\rm exp},j}|^2=|a_p^j|^2$ exactly. 
Thus, the random variables $|a_p^{{\rm exp},j}|^2$ are independent and follow  exponential distributions with mean $n_p^{\rm RJ}$.
Consequently, the empirical quantities $Z_p^Q= {n}_p^{{\rm exp},Q} / n_p^{\rm RJ}$ are independent and identically distributed with the gamma probability distribution $\Gamma(Q,Q)$ (the law of the sum of  $Q$ independent variables with exponential distribution and mean $1/Q$) and the estimation error is 
\begin{equation}
{\cal D}_{\rm err}^Q = \frac{\sum_p n_p^{\rm RJ} |Z_p^Q-1|}{\sum_p n_p^{\rm RJ} (Z_p^Q+1)} ,
\end{equation}
which gives when the number of modes is large enough ($Z^Q$ follows the $\Gamma(Q,Q)$ distribution):
\begin{equation}
{\cal D}_{\rm err}^Q \simeq \frac{\EE[|Z^Q-1|]}{\EE[Z^Q+1]} 
= \frac{Q^{Q-1}}{(Q-1)!} e^{-Q}  .
\label{eq:errQtheo}
\end{equation}
For $Q\geq 8$ , we have ${\cal D}_{\rm err}^Q \simeq 1/\sqrt{2\pi Q}$.

We have carried out numerical simulations with our implementation of the phase-retrieval algorithm (using multiple initial phase guesses) and with the sampling error of the camera. 
The results of the distance ${\cal D}_{\rm err}^Q$ vs energy $E$ are reported in Fig.~\ref{fig:D_vs_E} with different numbers $Q$ of realizations per energy. We can see in panel (b) of Fig.~\ref{fig:D_vs_E} that the errors  correspond to the theoretical error Eq.(\ref{eq:errQtheo}) when the phase-retrieval algorithm makes no error.

The error introduced by the Gerchberg-Saxton algorithm has been computed by increasing the amount of complexity in the speckle pattern, i.e., by increasing the energy $E$. We can see in Fig.~\ref{fig:D_vs_E}(a) that the error does not increase when the energy $E$ increases.

In the experiments we have typically $35$ to $70$ independent realizations of speckle beams for a given small energy interval $[E - \Delta E, E +\Delta E]$ with $\Delta E = 0.125E_{\rm min}$, so we can expect that the errors (due to the phase retrieval algorithm and the camera sampling) are small.
Error bars with relative standard deviations of the order of $1/\sqrt{2\pi Q} \simeq 6\%$ could be added in Fig.~2  but they are too small to be visible.

To complete our study, we report in Fig.~\ref{fig:2D} the near-field and far-field intensity distributions recorded during one of the experiments (left plots) and the corresponding reconstructed intensities from the Gerchberg-Saxton algorithm (right plots).


\subsection{Experimental convergence to the NT RJ distribution}
 
\begin{figure}
\includegraphics[width=1\columnwidth]{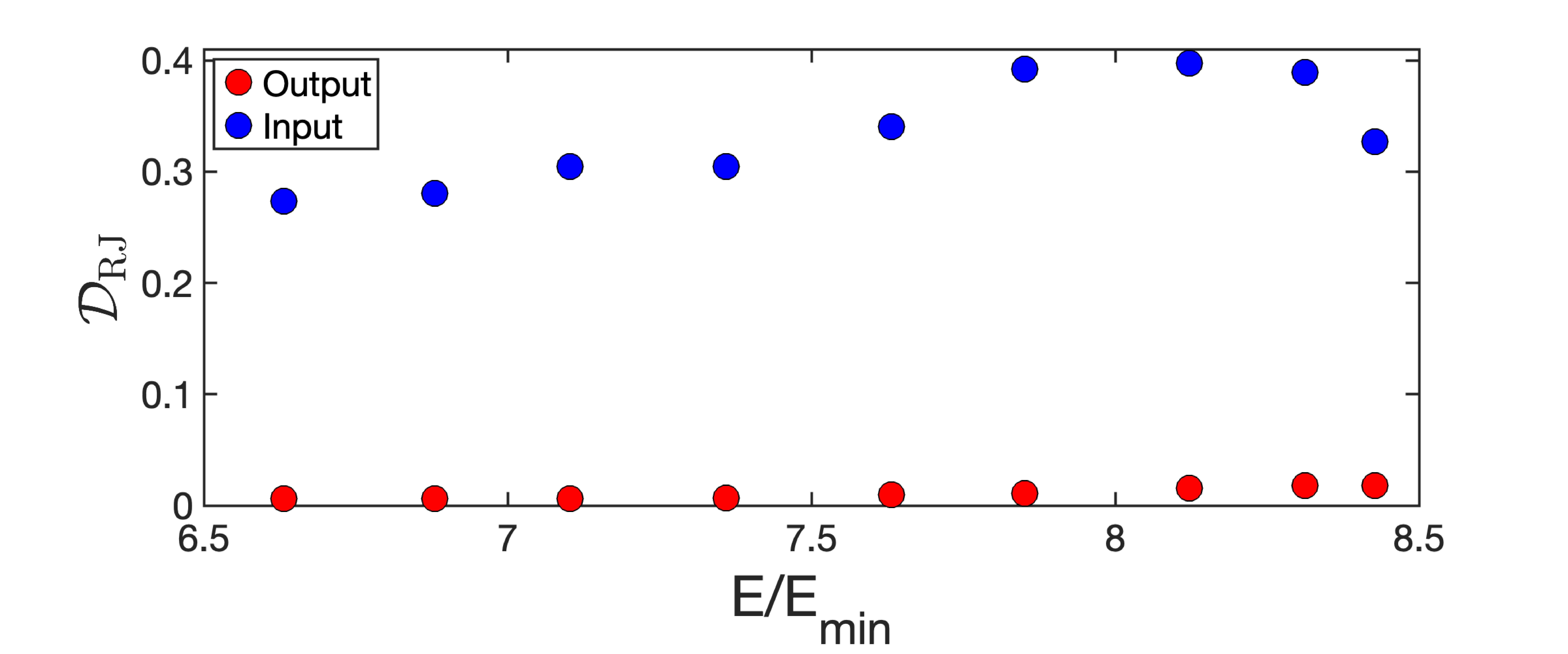}
\caption{
\baselineskip 10pt
{\bf Experimental attraction to NT RJ equilibrium.} 
Distance ${\cal D}_{\rm RJ}$ [defined in Eq.(\ref{eq:distance})] to the RJ equilibrium distribution computed from the experimental data at the fiber input (blue), and fiber output (red), for different values of the energy $E$.
The significant reduction of ${\cal D}_{\rm RJ}$ from input to output measurements shows the attraction to the RJ equilibrium for $T<0$.
Note that ${\cal D}_{\rm RJ}$ in Eq.(\ref{eq:distance}) is bounded, $0 \le {\cal D}_{\rm RJ} \le 1$.
We recall that negative temperatures $T<0$ occur for $E>E_*$ with $E_*/E_{\rm min}\simeq 6.33$, see Fig.~1.
}
\label{fig:D} 
\end{figure}

We have quantified in our experimental results the attraction to the NT equilibrium by using Eq.(\ref{eq:distance_sampl}), which provides a `distance' to the RJ distribution 
\begin{equation}
{\cal D}_{\rm RJ}=\frac{ \sum_p |n_p^{\rm exp} - n_p^{\rm RJ}| }{ \sum_p n_p^{\rm exp} + n_p^{\rm RJ}}.
\label{eq:distance}
\end{equation}
We report in Fig.~\ref{fig:D} the distance ${\cal D}_{\rm RJ}$ computed for the experimental data averaged over the realizations $n_p^{\rm exp}$ at the fiber input (blue), and the fiber output (red), for different energies $E$.
The strong reduction of the distance ${\cal D}_{\rm RJ}$ from the fiber input to the output confirms the process of NT thermalization, which is demonstrated over a broad range of values of the energy $E$. 

\subsection{Polarization effects}
 
\begin{figure}
\includegraphics[width=1\columnwidth]{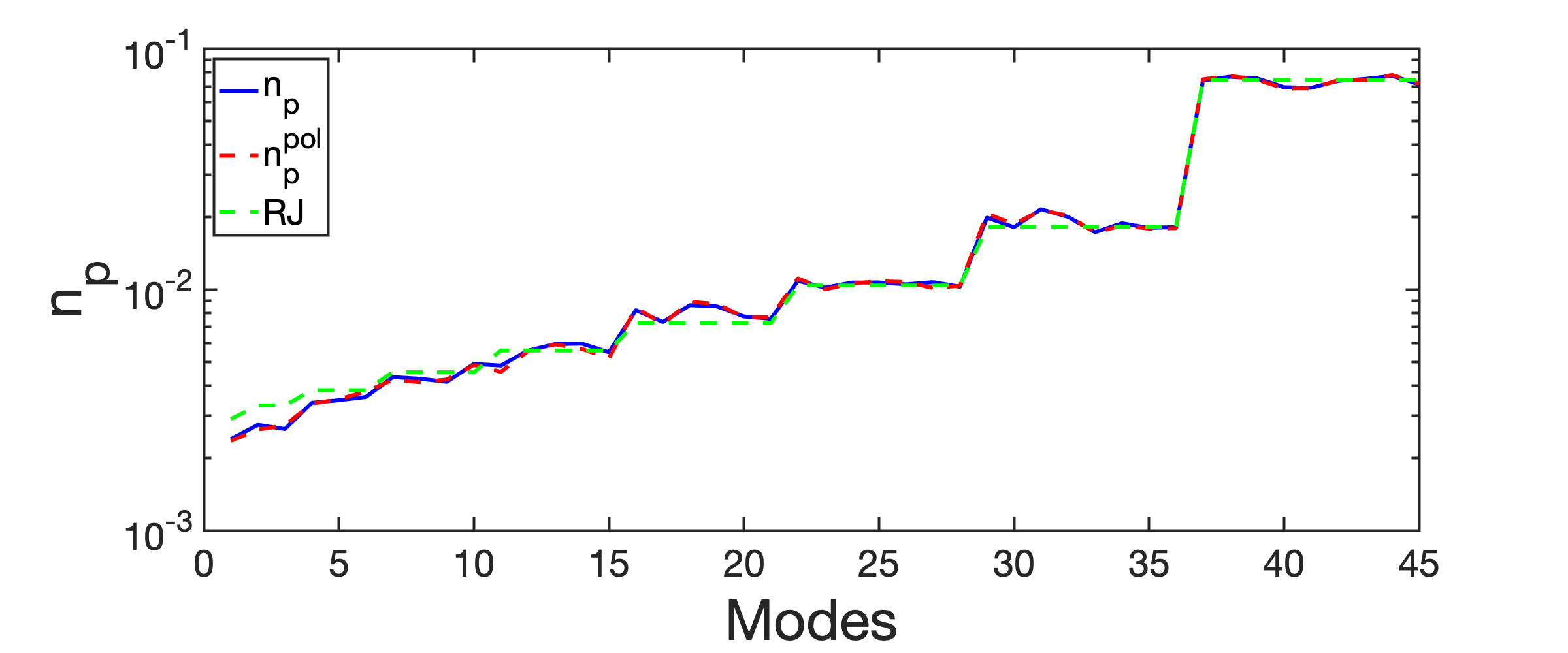}
\caption{
\baselineskip 10pt
{\bf Polarization effects in modal decomposition.} 
Experimental modal distribution retrieved from the Gerchberg-Saxton algorithm, without splitting the orthogonal polarization (blue line), by splitting the orthogonal polarization at the fiber output (dashed red line).
Note that a single realization of the speckle beam has been considered, which explains the discrepancy with the RJ equilibrium distribution (dashed green line).
}
\label{fig:polar} 
\end{figure}

The polarization state of the optical beam changes as it propagates through the MMF. 
The field at the output of the MMF is projected onto a basis of orthogonal linear polarizations. 
The corresponding NF and FF intensity distributions are recorded along the orthogonal linear polarizations.
For each polarization, we apply the mode decomposition procedure based on the Gerchberg-Saxton presented above.
In this way, we retrieve the transverse phase profile of the field for the two orthogonal polarizations.
The complex field along each polarization is subsequently projected over the fiber modes, to get the modal populations for each polarization, $n_p^{(x)}$ and $n_p^{(y)}$.
The modal distribution is obtained by summing the contributions of the two polarizations, $n_p^{\rm pol}=(N_x n_p^{(x)}+ N_y n_p^{(y)})/(N_x+N_y)$, where $N_{x,y}$ denote the power along the two polarizations.
We compare in Fig.~\ref{fig:polar} the modal distribution retrieved by following this procedure ($n_p^{\rm pol}$), with the modal distribution ($n_p$) retrieved without separating the polarization states of the field.
This comparison is reported in Fig.~\ref{fig:polar} for the same (single) realization of the speckle beam.
We can remark in Fig.~\ref{fig:polar} that the two distributions are almost identical.
In all cases analyzed, we have always observed the same good agreement.
Given the large number of realizations of speckle beams ($\simeq$300) recorded and analyzed in our experiments, we didn't perform the polarization modal decomposition.

\section{Analogue condensation at NT}

\subsection{Thermodynamics relations}

We define the thermodynamic relations used to plot Fig.~1.
We consider a field at equilibrium with the RJ distribution $n_p^{\rm RJ}=T/(\beta_p - \mu)$, with $N=\sum_p n_p^{\rm RJ}$ and $E= \sum_p \beta_p n_p^{\rm RJ}$, 
where $\beta_p=\beta_{p_x,p_y}=\beta_0(p_x+p_y+1)$ are the eigenvalues of the truncated parabolic potential ($\beta_p \le \beta_{\rm max}$), and the index $\{p\}$ labels the two integers $(p_x,p_y)$ that specify a mode.
Here and below, the sum over the modes $\sum_p$ is carried over the set $0\le p_x+p_y < g$,
where $g=\beta_{\rm max}/\beta_0$ is the number of groups of non-degenerate modes, with $M=g(g+1)/2$ the total number of modes.
Because of the constraint $n_p^{\rm RJ}=T/(\beta_p - \mu) >0$, a NT equilibrium state $T<0$ requires that $\mu > \beta_{\rm max}$.

We start from the equilibrium entropy ${\tilde S}^{eq}=\sum_p \log(n_p^{eq})$ -- note that at equilibrium it coincides with the nonequilibrium entropy verifying the $H$ theorem of entropy growth. 
It proves convenient to shift the entropy by a constant ${S}^{eq}={\tilde S}^{eq}- M \log N$, so that by using  $T=N/\sum_p(\beta_p-\mu)^{-1}$, we can write
\begin{eqnarray}
{S}(\mu)&=& - \sum_p\log ( \beta_p-\mu ) -M \log\Big( \sum_p \frac{1}{\beta_p-\mu} \Big) 
\label{eq:S_mu}\\
\frac{E(\mu)}{E_{\rm min}} &=& \frac{ \sum_p \frac{\beta_p}{\beta_p-\mu}  }{ \sum_p \frac{\beta_0}{\beta_p-\mu}  }
\label{eq:E_mu}\\
\frac{T(\mu)}{E_{\rm min}} &=& \frac{1}{\sum_p \frac{\beta_0}{\beta_p-\mu}}
\label{eq:T_mu}
\end{eqnarray}
The parametric plot with resp. to $\mu$ of (\ref{eq:S_mu}) and (\ref{eq:E_mu}) gives ${S}(E)$ in Fig.~1(b); the corresponding parametric plot of (\ref{eq:E_mu}) and (\ref{eq:T_mu}) gives $T$ vs $E$ in Fig.~1(c).

\begin{figure}
\includegraphics[height=8.25cm,width=8.25cm]{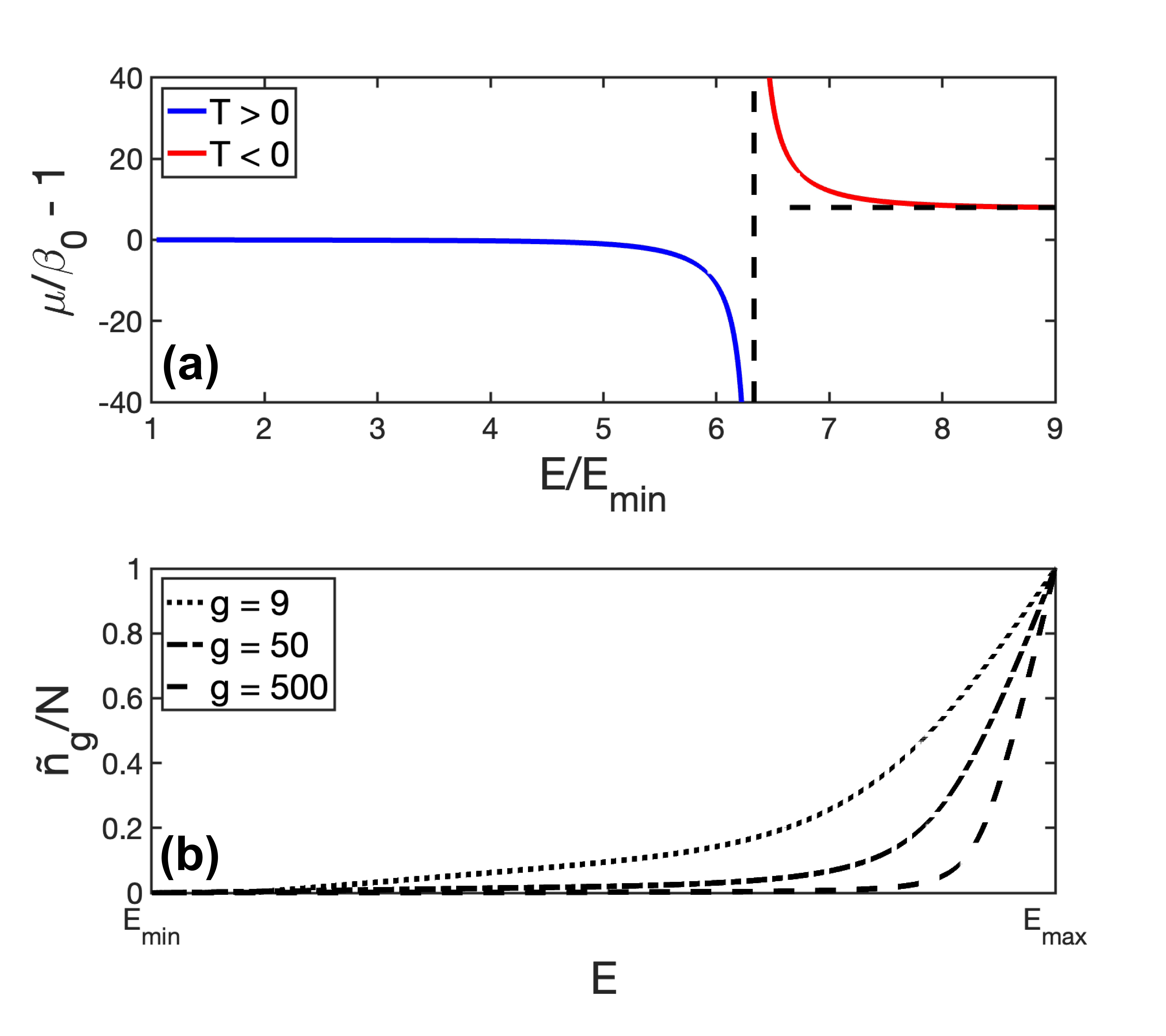}
\caption{
\baselineskip 10pt
{\bf Analogue effect of condensation at NTs:} 
(a) Chemical potential $\mu/\beta_0 - 1$ vs energy $E/E_{\rm min}$ for the MMF used in the experiments ($g=9$), from Eq.(\ref{eq:E_mu}). Note the asymptotic behaviors $\mu \to \beta_0^-$ for $E \to E_{\rm min}$, and $\mu \to \beta_{\rm max}^+$ for $E \to E_{\rm max}$, which lead respectively to the macroscopic population of the lowest energy level ($\beta_0$), and highest energy level ($\beta_{\rm max}$).
The horizontal dashed line denotes $\mu=\beta_{\rm max}$ and the vertical one $E=E_*$.
(b) Condensate fraction in the highest energy level ${\tilde n}_g^{\rm RJ}/N$ vs energy $E$: As the energy increases $E \to E_{\rm max}$ (or equivalently the temperature increases $T \to 0^-$), the condensate fraction increases ${\tilde n}_g^{\rm RJ}/N  \to 1$.
The curves are obtained from Eqs.(\ref{eq:n_0RJ_neg}-\ref{eq:E_mu_neg}), see the text. 
The critical behavior of the transition to condensation becomes apparent by increasing the number of modes $g$.
}
\label{fig:NT_cond} 
\end{figure}


\subsection{NT states in the thermodynamic limit}

The thermodynamic limit is defined by $N \to \infty$, $\beta_0 \to 0$ with $N \beta_0^2=$const and $\beta_{\rm max}=$const. 
In this limit the discrete sums over the modes are replaced by continuous integrals, namely $N=\sum_p n_p^{eq} \to (T/\beta_0^2) \int_0^{\beta_{\rm max}} dx \int_0^{\beta_{\rm max}-x} dy (x+y+\beta_0-\mu)^{-1}$, 
which gives
\begin{eqnarray}
N \beta_0^2 = T \beta_{\rm max} \Big( 1 - z  \log\big( z/(z-1 )\big)   \Big),
\label{eq:N_TL}
\end{eqnarray}
where $z=\mu/\beta_{\rm max}$.
A negative temperature equilibrium state ($T<0$) is characterized by $z > 1$.
It can exist in the thermodynamic limit with the constraint $N \beta_0^2 = {\rm const} >0$ provided that  $1 - z  \log\big( z/(z-1 )\big) < 0$.
This inequality is always verified for $z>1$, which confirms the existence of NT equilibrium states in the thermodynamic limit.


\subsection{NT condensation in the highest energy level}

\noindent 
{\it Positive temperature $T>0$.} We start by briefly summarizing the usual positive temperature condensation in the lowest energy level. This effect of condensation originates in the {\it singularity} of the RJ distribution. Indeed, the denominator of $n_p^{RJ}=T/(\beta_p-\mu)$ vanishes for the lowest energy level when $\mu = \beta_0$. This gives rise to an analogue effect of condensation: As the energy decreases below a critical value $E_{\rm crit}$ (or $T < T_{\rm crit}$), then $\mu \to \beta_0^-$ (see Fig.~\ref{fig:NT_cond}(a)) and the singular behavior of the RJ distribution is regularized by the macroscopic population of the fundamental mode: 
\begin{eqnarray}
n_0^{\rm RJ}/N \to 1 \quad {\rm as} \quad E \to E_{\rm min} \ \ ({\rm or} \ T \to 0^+).
\end{eqnarray}
It has been shown that this condensation-like effect is a phase transition that occurs in the thermodynamic limit, see Ref.\cite{PRL20}.

\noindent  \smallskip 

{\it Negative temperature, $T<0$.} In the NT region, we reveal an inverted condensation-like effect, which is characterized by a macroscopic population of the highest energy level.
Aside from the singularity for $\mu=\beta_0$ discussed here above for $T>0$, the RJ distribution $n_p^{RJ}=T/(\beta_p-\mu)$ also exhibits a singularity (vanishing denominator) for the highest energy level when $\mu = \beta_{\rm max}$ (see Fig.~\ref{fig:NT_cond}(a)). 
We have shown in Fig.~5, that the highest energy level $g=9$ becomes macroscopically populated as the energy increases:
\begin{eqnarray}
{\tilde n}_g/N \to 1 \quad  {\rm as} \quad E \to E_{\rm max} \ \ ({\rm or} \ T \to 0^-),
\end{eqnarray}
with $\tilde{n}_g = \sum_{p, p_x+p_y+1=9} n_p$. 
This condensation-like effect does not occur in the thermodynamic limit.
We pose $\mu = \beta_{\rm max} + \varepsilon$, with $\varepsilon > 0$.
Eq.(\ref{eq:N_TL}) can be written in the limit $\varepsilon \to 0^+$: 
$N \beta_0^2 \simeq T \beta_{\rm max} \big( 1 - \log( \beta_{\rm max}/\varepsilon )   \big)$.
By keeping $N \beta_0^2 =$const, the chemical potential $\mu$ reaches $\beta_{\rm max}^+$ for a vanishing temperature $T \to 0^-$, i.e., condensation does not occur in the thermodynamic limit.
However, an analogue effect of condensation occurs through a macroscopic population of the highest energy level. 
By setting $\beta_p=\beta_{\rm max}$ in the RJ distribution, then $n_g^{\rm RJ}=-T/(\mu-\beta_{\rm max})$ denotes the power in one mode of the highest energy level.
The total power in the highest energy level ($g-$fold degenerate) ${\tilde n}_g^{\rm RJ}=g n_g^{\rm RJ}$  then reads
\begin{eqnarray}
\frac{{\tilde n}_g^{\rm RJ}}{N}(\mu)  & =&   \frac{g}{(\mu-\beta_{\rm max}) \sum_p (\mu-\beta_p)^{-1}},
\label{eq:n_0RJ_neg}\\
\frac{E(\mu)}{E_{\rm min}} &=& \frac{ \sum_p \frac{\beta_p}{\beta_p - \mu}  }{\sum_p \frac{\beta_0}{\beta_p - \mu}  }.
\label{eq:E_mu_neg}
\end{eqnarray}
The parametric plot of (\ref{eq:n_0RJ_neg}) and (\ref{eq:E_mu_neg}) with respect to $\mu$ provides the condensate fraction reported in Fig.~5.
Fig.~\ref{fig:NT_cond}(b) shows the condensate fraction, ${\tilde n}_g^{\rm RJ}/N$ vs $E$, by increasing the number of modes, i.e., by approaching the thermodynamic limit.
The critical behavior of the condensation curve looks similar to that of a  phase transition, thought strictly speaking phase transitions only occur in the thermodynamic limit.
Nevertheless, if one takes the macroscopic occupation of an energy level as the essential characteristic of condensation, then NTs are characterized by a condensation-like effect into the highest-energy level.

\begin{figure}
\includegraphics[width=1\columnwidth]{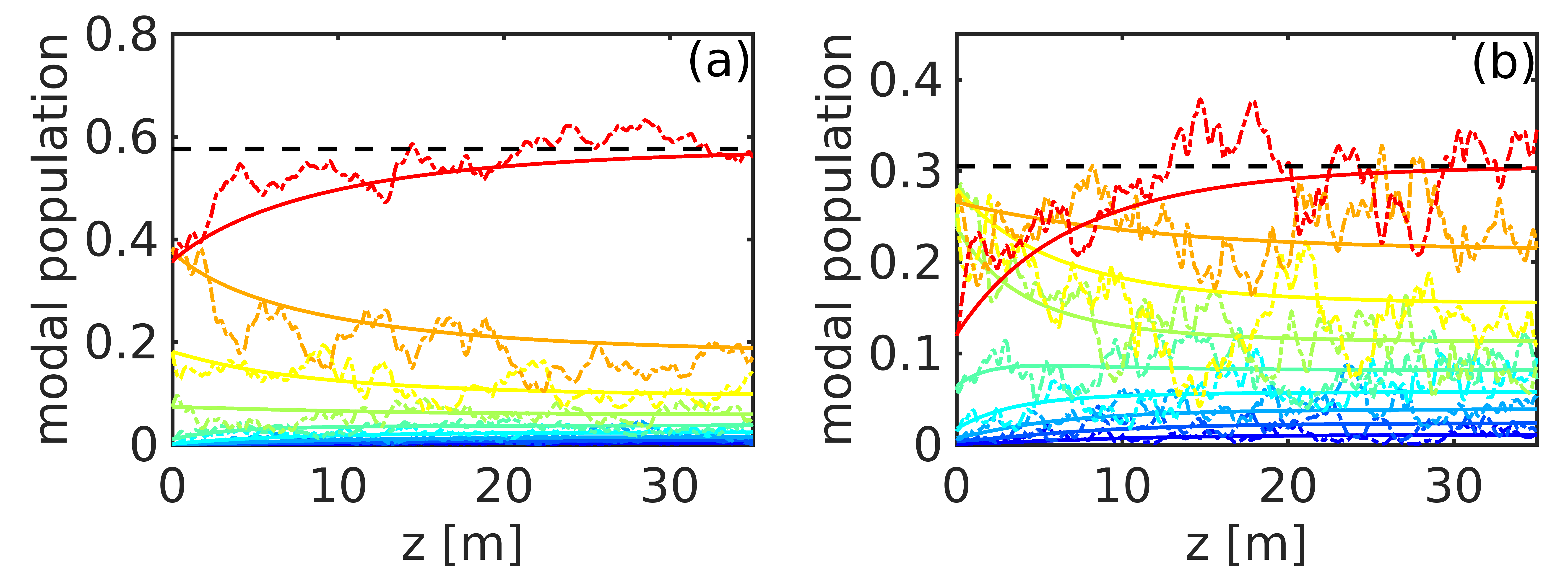}
\caption{
\baselineskip 10pt
{\bf Simulations of NLS equation and wave turbulence kinetic equation:} 
Simulations of the NLS equation (dashed lines), and wave turbulence kinetic equation (solid lines) showing the evolutions during the propagtion (in $z$) of the power within each of the $g=$9 groups of degenerate modes of the fiber that has been used in the experiments, for $E/E_{\rm min}=8$ (a), $E/E_{\rm min}=7$ (b). 
The horizontal dashed black line denotes the population for the higher mode group at complete RJ thermal equilibrium.
See the text for details on initial conditions and parameters.
}
\label{fig:simul} 
\end{figure}

\noindent \smallskip 

\section{Wave turbulence simulations}

The starting point is the 
NLS equation governing light propagation in MMFs \cite{mumtaz13,PRA19}.
By expanding the random wave into the fiber modes and considering the dominant contribution of weak disorder, the polarized modal components ${\bm a}_p=(a_{p,x},a_{p,y})^T$ are governed by \cite{PRA19}:
\begin{eqnarray}
i \partial_z {\bm a}_p = \beta_p {\bm a}_p + {\bf D}_p(z) {\bm a}_p - \gamma  {\bm P}_p({\bm a}),
\label{eq:nls_Ap}
\end{eqnarray}
where 
${\bm P}_p ({\bm a})  = \sum_{l,m,n} S_{plmn} \Big(\frac{1}{3}
{\bm a}_l^T {\bm a}_m {\bm a}_n^* +\frac{2}{3} {\bm a}_n^\dag  {\bm a}_m {\bm a}_l\Big)$,
$S_{plmn}$ 
denoting the spatial overlap among the modes.
The introduction of disorder is important in order to avoid strong phase correlations among the modes that are related to Fermi-Pasta-Ulam recurrences \cite{malomed21}, which inhibit the thermalization process \cite{PRL19}.
We consider the most general form of disorder that conserves the power: 
The Hermitian matrices ${\bf D}_p(z)$ are expanded into the Pauli matrices that form a basis for the vector space of 2$\times$2 Hermitian matrices.
The matrices then have the form
${\bf D}_p(z) = \sum_{j=0}^3 \nu_{p,j} (z) \bsigma_j$,
where $\bsigma_j$ ($j=1,2,3$) are the Pauli matrices ($\bsigma_0$ the identity matrix), while
$\nu_{p,j}(z)$ are independent and identically distributed real-valued random processes, with variance $\sigma^2$ and correlation length $\ell_c$.
The corresponding characteristic length scale of disorder is $L_{d}=1/\Delta \beta$, with $\Delta \beta = \sigma^2 \ell_c$, see Ref.\cite{PRA19} for details.

The nonequilibrium process of NT thermalization can be described by a wave-turbulence kinetic equation.
The kinetic equation was derived from the modal NLS Eq.(\ref{eq:nls_Ap}) in the weakly nonlinear regime of the experiment, $L_{lin}\sim 1/\beta_0 \ll L_{nl} \sim 1/(\gamma N)$. 
It describes the nonequilibrium evolution of the averaged modal components $n_p(z)=\left< |{\bm a}_p|^2(z)\right>$ \cite{PRA19}:
\begin{eqnarray}
\nonumber
\partial_z n_p(z) &=&  \frac{ \gamma^2}{6\Delta \beta} \sum_{l,m,n} |S_{lmnp}|^2 \delta^K(\Delta\omega_{lmnp}) M_{lmnp}({\bm n}) \\
&&
+  \, \frac{4\gamma^2}{9 \Delta \beta}  \sum_l  
 |  s_{lp}({\bm n}) |^2 \delta^K(\Delta\omega_{lp})   (n_l-n_p), \quad \quad
\label{eq:kin}
\end{eqnarray}
with $s_{lp}({\bm n})=\sum_{m'} S_{lm'm'p} n_{m'}$, and $M_{lmnp}({\bm n})=  n_l n_m n_p+n_l n_m n_n -  n_n n_p n_m -n_n n_p n_l$ and $\Delta \omega_{lp}=\beta_l-\beta_p$.
The term $\delta^K(\Delta\omega_{lmnp})$ denotes the four-wave frequency resonance 
$\Delta \omega_{lmnp}  = \beta_l+\beta_m-\beta_n-\beta_p$, with 
$\delta^K(\Delta \omega_{lmnp})=1$ if $\Delta \omega_{lmnp}=0$, and zero otherwise.
The kinetic Eq.(\ref{eq:kin}) conserves $N=\sum_p n_p(z)$, the energy $E=\sum_p \beta_p n_p(z)$ and exhibits a $H-$theorem of entropy growth, $\partial_z {\cal S}_{\rm kin}(z) \ge 0$, for the nonequilibrium entropy ${\cal S}_{\rm kin}(z)=\sum_p \log\big(n_p(z)\big)$. Accordingly, it describes an irreversible evolution of the speckle beam to the RJ equilibrium distribution realizing the maximum of entropy, $n^{\rm RJ}_p=T/(\beta_p - \mu)$ \cite{PRA19}.

We have considered in the simulations the MMF used in the experiments, see Sec.~\ref{sec:setup}.
The initial condition consists of a speckle beam whose correlation length is varied in such a way to fix a desired value of the energy $E$.
The considered parameters are $\ell_c=0.3$m and $2\pi/\sigma=2.1$m \cite{PRA19}. 
The results of the simulations of the NLS equation and the corresponding simulations of the wave turbulence kinetic equation are reported in Fig.~\ref{fig:simul}.
They show the process of thermalization to the negative temperature RJ equilibrium state predicted by the theory.
The simulations qualitatively reproduce the experimental results, although a power of 20kW has been considered to accelerate the dynamics.
The purely spatial model considered here then captures the essential features of the condensation process reported experimentally. 

At variance with NLS simulations that are stochastic and thus exhibit fluctuations, the simulations of the wave turbulence kinetic equation are deterministic (free of fluctuations).
A good agreement between NLS  and kinetic simulations is obtained without using any adjustable parameter.
The wave turbulence kinetic equation then provides a nonequilibrium description of the process of negative temperature thermalization observed experimentally.



\begin{thebibliography}{99}
\bibitem{comment}
Note that, particles in a laser have an inverted energy population which, however, only exists as long as the laser is continuously pumped (by switching off the pump, all atoms decay back into the lower state and their energy goes into the laser beam). The laser medium is in a steady state, but not in thermal equilibrium and therefore cannot have a temperature.
\end{thebibliography}

\end{document}